\documentclass[11pt] {article}

\begin{document}
\title
{\bf Subsystem of neutral mesons\\ beyond the Lee--Oehme--Yang
approximation}
\author{K. Urbanowski\footnote{e--mail:
K.Urbanowski@if.uz.zgora.pl; K.Urbanowski@proton.if.uz.zgora.pl}
 \\ University of Zielona Gora, Institute of Physics,\\ ul. Podgorna
50, 65--246 Zielona Gora, Poland.} \maketitle

\begin{abstract}
We  begin with a discussion of the general properties of
eigenvectors and eigenvalues for an effective Hamiltonian
governing time evolution in a two state subspace of the state
space of the total system under consideration. Next, the Lee,
Oehme and Yang (LOY) theory of  time evolution in such a subspace
is considered. The CPT-- and CP--symmetry properties of the LOY
effective Hamiltonian are discussed. Next the CPT transformation
properties of the exact effective Hamiltonian for two state
subspace are discussed. Using the Khalfin Theorem we show that the
diagonal matrix elements of the exact effective Hamiltonian
governing the time evolution in the subspace of states of an
unstable particle and its antiparticle need not be equal at for $t
> t_{0}$ ($t_{0}$ is the instant of creation of the pair) when the
total system under consideration is CPT invariant but CP
noninvariant. (Suitable matrix elements of the LOY effective
Hamiltonian are equal in such a case). The unusual consequence of
this result is that, contrary to the properties of stable
particles, the masses of the unstable particle "1" and its
antiparticle "2" need not be equal for $t \gg t_{0}$ in the case
of preserved CPT and violated CP symmetries. Also, basic
assumptions necessary for the proof of the  CPT Theorem are
discussed. It is found that the CPT Theorem is not valid for a
physical system with unstable particles decaying exponentially.
From this property the conclusion is drawn that
CPT--transformation cannot be a symmetry in a system which
contains the LOY model as a subsystem, and, thus this model is
shown to be incapable of describing possible CPT--violation
effects correctly. Using an exact equation governing the time
evolution in the subspace of the total state space we show that
there exists an approximation which is more accurate than the LOY
approximation, and which leads to an effective Hamiltonian whose
diagonal matrix elements posses properties consistent with the
conclusions obtained for the exact effective Hamiltonian. Using
this more accurate approximation  we show that the interpretation
of the tests measuring the difference between the $K_{0}$ mass and
the ${\overline K}_{0}$ mass as the CPT--symmetry test is wrong.
We find that in fact such tests should rather be considered as
tests for the existence of the hypothetical interaction allowing
the first order $|\Delta S| = 2$ transitions $K_{0}
\rightleftharpoons {\overline K}_{0}$. We also discuss relations
between some parameters describing properties of the neutral meson
complex obtained within the LOY theory and those obtained using
the more accurate approximation than the LOY one.
\end{abstract}

\section[1]{Introduction.}

Many   tests   of fundamental symmetries consist in searching  for
decay processes of neutral mesons. A subsystem of neutral mesons
forms a two particle complex in the total system under
investigations. In the quantum decay theory of multiparticle
complexes, properties of the transition amplitudes
\begin{equation}
A_{j, \psi }(t) = \langle u_{j}| \psi ;t\rangle  \label{urb-amp}
\end{equation}
and properties of the matrix elements of the effective Hamiltonian
governing the time evolution in the subspace of states of the
complex are usually analyzed. Here, vectors $\{ |u_{j}\rangle
{\}}_{j \in U} \in {\cal H}$, where $\cal H$ is the total state
space of the system under consideration, represent the unstable
states of the system considered, $\langle u_{j}|u_{k}\rangle  =
{\delta}_{jk}$, and $| \psi ; t \rangle  $ is the solution of the
Schr\"{o}dinger equation (we use $\hbar = c = 1$ units)
\begin{equation}
i \frac{\partial}{\partial t} |{\psi};t\rangle   = H
|{\psi};t\rangle , \label{urb-Schrod}
\end{equation}
having the following form
\begin{equation}
| \psi ;t> = \sum_{j \in U}a_{j}(t) |u_{j}\rangle  + \sum_{J}
f_{J}(t)| {\phi}_{J}\rangle  , \label{urb-psi-gen}
\end{equation}
where vectors $|{\phi}_{J}\rangle \in {\cal H}$ describe the
states of decay products, $\langle u_{j}|{\phi}_{J}\rangle  =0$
for every $j \in U$. The initial condition for Eq
(\ref{urb-Schrod}) in the case considered is usually assumed to be
\begin{eqnarray}
| \psi ;t = t_{0} \equiv 0 \rangle  & \stackrel{\rm def}{=} & |
\psi \rangle  \equiv \sum_{j \in U} a_{j} |u_{j}\rangle \in {\cal
H}_{||}, \label{urb-init0} \\
f_{J} (t = 0) & = & 0. \nonumber
\end{eqnarray}
where ${\cal H}_{||}$ is the subspace of $\cal H$ spanned by set
of vectors $\{ |u_{j}\rangle {\}}_{j \in U}$. In Eq
(\ref{urb-Schrod})  $H$ denotes the complete (full), selfadjoint
Hamiltonian of the system.

From (\ref{urb-Schrod}) it follows that in the general case $| \psi
; t\rangle  = U(t) |\psi \rangle $, where
\begin{equation}
U(t) \equiv e^{-itH}, \label{urb-U(t)}
\end{equation}
is the total unitary evolution operator. So,
\begin{equation}
A_{j,\psi}(t) \equiv \langle u_{j}|U(t)|\psi \rangle  .
\label{urb-A-jk}
\end{equation}
It is not difficult to see that this property and hermiticity of
$H$ imply that
\begin{equation}
A_{j,j}(t)^{\ast} =A_{j,j}(-t) . \label{urb-amp-ast}
\end{equation}
Therefore, the decay probability of an unstable state (usually
called the decay law), i.e., the probability for a quantum system
to remain in its initial state $| \psi \rangle  \equiv
|u_{j}\rangle $
\begin{equation}
p_{j} (t) \stackrel{\rm def}{=} |A_{j,j}(t)|^{2} \equiv
|a_{j}(t)|^{2}, \label{urb-prob}
\end{equation}
must be an even function of time:
\begin{equation}
p_{j} (t) = p_{j} ( - t). \label{urb-even}
\end{equation}

This last property suggests that in the case of the unstable
states prepared at some instant $t_{0}$, say $t_{0} = 0$,  the
initial condition (\ref{urb-init0}) for the evolution equation
(\ref{urb-Schrod}) should be formulated more precisely. Namely,
from (\ref{urb-even}) it follows that the probabilities of finding
the system in the decaying state $|u_{j}\rangle $ at the instant,
say $t=T \gg t_{0} \equiv 0$, and at the instant $t =-T$ are the
same. Of course, this can never occur. In almost all experiments
in which the decay law of a given unstable particle is
investigated this particle is created at some instant of time, say
$t_{0}$, and this instant of time is usually considered as the
initial instant for the problem. From the property
(\ref{urb-even}) it follows that the instantaneous creation  of
the unstable particle is impossible. For the observer, the
creation of this particle (i.e., the preparation of the state,
$|u_{j}\rangle $, representing the decaying particle) is
practically instantaneous. What is more, using suitable detectors
he is usually able to prove that it did not exist at times $t <
t_{0}$.  Therefore, if one looks for the solutions of the
Schr\"{o}dinger equation (\ref{urb-Schrod}) describing properties
of the unstable states prepared at some initial instant $t_{0}$ in
the system, and if one requires these solutions to reflect
situations  described above, one should complete initial
conditions (\ref{urb-init0}) for Eq (\ref{urb-Schrod}) by assuming
additionally that
\begin{equation}
a_{j}(t < t_{0}) = 0, \; \; ( j \in U), \label{urb-init01}
\end{equation}
and that, for the problem, time $t$ varies from $t = t_{0} > -
\infty$ to $t = + \infty$ only.

Amplitudes of type $a_{j}(t)$ can be calculated directly by
solving the evolution equation (\ref{urb-Schrod}),  or by using
the Schr\"{o}dinger--like evolution equation governing the time
evolution in ${\cal H}_{||}$. Searching for the properties of two
particle subsystems one usually uses the following equation of the
type mentioned \cite{LOY1} --- \cite{Bigi} instead of Eq
(\ref{urb-Schrod}),
\begin{equation}
i \frac{\partial}{\partial t} |\psi ; t \rangle _{\parallel} =
H_{\parallel} |\psi ; t \rangle _{\parallel}, \label{urb-H-par}
\end{equation}
where $|\psi ; t\rangle_{||} \in {\cal H}_{||}$ and by
$H_{\parallel}$ we denote the effective nonhermitian Hamiltonian,
which in general can depend on time $t$ \cite{horwitz},
\begin{equation}
H_{\parallel} \equiv M - \frac{i}{2} \Gamma, \label{urb-H-par0}
\end{equation}
and
\begin{equation}
M = M^{+}, \; \; \Gamma = {\Gamma}^{+}, \label{urb-M-G}
\end{equation}
are $(2 \times 2)$ matrices, acting in a two--dimensional subspace
${\cal H}_{\parallel}$ of the total state space $\cal H$. $M$ is
called the mass matrix, $\Gamma$ is the decay matrix \cite{LOY1}
--- \cite{Bigi}. In many papers it is assumed that the real parts,
$\Re (.)$, of
the diagonal matrix elements of $H_{\parallel}$:
\begin{equation}
\Re \, (h_{jj} ) \equiv M_{jj} \stackrel{\rm def}{=} M_{j}, \;
\;(j =1,2), \label{urb-m-jj}
\end{equation}
where
\begin{equation}
h_{jk}  =  <{\bf j}|H_{\parallel}|{\bf k}>, \; (j,k=1,2),
\label{urb-h-jk}
\end{equation}
correspond to the masses, $M_{1},\, M_{2}$,  of particle "1" and
its antiparticle "2" respectively \cite{LOY1} --- \cite{Bigi},
(and such an interpretation of $\Re \, (h_{11})$ and $\Re \,
(h_{22})$ will be used in this paper), whereas the imaginary
parts, $\Im (.)$,
\begin{equation}
-2 \Im \, (h_{jj}) \equiv {\Gamma}_{jj} \stackrel{\rm def}{=}
\Gamma_{j}, \; \;(j =1,2), \label{urb-g-jj}
\end{equation}
are interpreted as the decay widths of these particles \cite{LOY1}
--- \cite{Bigi}.

The standard method of derivation of such a $H_{\parallel}$ is
based on a modification of the Weisskopf--Wigner (WW)
approximation \cite{ww}. Lee, Oehme and Yang (LOY) adapted the WW
approach to the case of a two particle subsystem \cite{LOY1}
--- \cite{comins} to obtain their effective Hamiltonian $H_{\parallel}
\equiv H_{LOY}$. Almost all properties of the neutral kaon
complex, or another two state subsystem, can be described by
solving Eq (\ref{urb-H-par}) \cite{LOY1}
--- \cite{Bigi}, with the initial condition corresponding to
(\ref{urb-init0}) and (\ref{urb-init01})
\begin{eqnarray}
| \psi ;t = t_{0} \rangle _{\parallel} & \equiv &
| \psi \rangle _{\parallel}, \nonumber \\
\parallel \, |{\psi} ; t = t_{0} \rangle _{\parallel}
{\parallel} & = & 1, \; \; | {\psi}; t < t_{0} \rangle
_{\parallel} = 0, \label{urb-init}
\end{eqnarray}
for $| \psi ; t \rangle _{\parallel}$ belonging to the subspace
${\cal H}_{\parallel} \subset {\cal H}$ spanned, e.g., by
orthonormal neutral  kaons states $|K_{0}\rangle , \;
|{\overline{K}}_{0}\rangle $, and so on, (then states
corresponding to the decay products belong to ${\cal H}
\ominus{\cal H}_{\parallel} \stackrel{\rm def}{=} {\cal
H}_{\perp}$),
\begin{equation}
|\psi \rangle _{\parallel} \equiv a_{1}|{\bf 1}\rangle  +
a_{2}|{\bf 2}\rangle , \label{urb-psi-par}
\end{equation}
and $|{\bf 1}\rangle $ stands for  the  vectors of the
$|K_{0}\rangle , \; |B_{0}\rangle $, etc., type and $|{\bf
2}\rangle $ denotes states of $|{\overline{K}}_{0}\rangle , \;
{\overline{B}}_{0}\rangle $ type, $\langle {\bf j}|{\bf k}\rangle
= {\delta}_{jk}$, $j,k =1,2$.

Defining projectors
\begin{equation}
P  =  |{\bf 1} \rangle  \langle {\bf 1}| + |{\bf 2}\rangle \langle
{\bf 2}|, \;\;\; Q = I - P, \label{urb-P}
\end{equation}
one has
\begin{equation}
{\cal H}_{||} = P {\cal H} \ni P |\psi ; t\rangle \stackrel{\rm
def}{=} |\psi ;t\rangle_{||}, \label{urb-psi-P}
\end{equation}
and
\begin{equation}
{\cal H}_{\perp} = Q {\cal H} \ni Q|\psi ; t \rangle \stackrel{\rm
def}{=} |\psi ; t \rangle_{\perp}. \label{urb-psi-Q}
\end{equation}
Solutions of Eq. (\ref{urb-H-par}) can be written  in a matrix form
and such a matrix  defines  the evolution    operator (which is
usually nonunitary) $U_{\parallel}(t)$ acting  in  ${\cal
H}_{\parallel}$:
\begin{equation}  |\psi  ;   t   \rangle _{\parallel} =
U_{\parallel}(t)  |\psi  ;t_{0}  =  0  \rangle _{\parallel}
\stackrel{\rm def}{=}   U_{\parallel}(t)    |\psi \rangle
_{\parallel}. \label{urb-U||-psi}
\end{equation}

The problem of testing such fundamental symmetries like parity,
CP--symmetry, or CPT--invariance, experimentally has attracted the
attention  of physicists,  practically  since the discovery of
antiparticles. An especially important problem seems to be the
problem of verifying if CPT symmetry is the symmetry of the
Nature. There is a known theorem (called  the CPT Theorem) in
axiomatic quantum field theory that CPT--invariance must hold.
This theorem is based on very general assumptions \cite{pauli} ---
\cite{wightman}: it requires for its proof that a  quantum field
theory be constructed   from fields belonging   to
finite--dimensional representations of the Lorentz group, have  a
local  interactions invariant  under  the proper  Lorentz  group,
and  the  spectral condition  be fulfilled (all  energies  must be
nonnegative). CPT--invariance is the minimal  condition  for the
existence  of antiparticles  within quantum  field  theory. Many
tests   of CPT--invariance consist in searching  for  decay
processes  of neutral kaons. It was realized almost from  the
discovery  of  CP violation \cite{cronin-1964} that it  was
important  to study  in  detail the effective Hamiltonian,
$H_{||}$, which describes the time evolution  in the $K_{0},
{\overline K}_{0}$ complex \cite{LOY1} --- \cite{Bigi}. In  the
large literature, all CP-- and CPT--violation parameters in this
complex are expressed in terms of matrix elements  of this
Hamiltonian. The  old and  new \cite{cronin} --- \cite{dafne},
\cite{barmin,lavoura},\cite{Jarlskog} --- \cite{Bigi} experimental
tests of such fundamental symmetries as the CP--nonivariance and
of the CPT--invariance in the neutral mesons complex need a
correct interpretation   of the parameters measured, which is
independent of the approximations used  for the theoretical
description of the behavior of such a complex. The aim of this
paper is to confront the standard interpretation of these
parameters with the general, basic principles of quantum theory.

The paper is organized as follows: We begin from the discussion of
Lee, Oehme and Yang theory, then we confront CP-- and
CPT--transformation properties of the LOY effective Hamiltonian
with corresponding properties of the exact effective Hamiltonian
for two particle complex under consideration. We also confront
assumptions of the Lee, Oehme and Yang theory with  assumptions
necessary  for  the proof  of  the CPT Theorem. Also, an
alternative approximation, (different  from that used by Lee,
Oehme and Yang), leading to the effective Hamiltonian, which CP--
and CPT--transformation properties are consistent  with such the
properties  of  the Hamiltonian  of the total  system considered,
is presented.

\section{Preliminaries}

\subsection{The  eigenvalue  problem in two--dimensional \\
subspace.}

In  the general case, the eigenvectors for $H_{||}$ are identified
with quasistationary states in the subspace of states, ${\cal
H}_{\parallel}$, of neutral meson complex under studies. When
studying the eigenvalue problem for the $(2 \times  2)$ matrices,
say ${\bf E}$, acting in ${\cal H}_{\parallel}$, one must solve
the equation
\begin{equation}
\det |{\bf E} - \zeta  I_{\parallel}| = 0 . \label{urb-E-eigen}
\end{equation}
Here $I_{\parallel}$ is the unity in ${\cal  H}_{\parallel}$  - of
course, $I_{\parallel} \equiv  P$. Sometimes it is useful to
rewrite the matrix ${\bf E}$ in terms of the Pauli matrices. In
such a case we have
\begin{equation}
{\bf E} = E_{0} I_{||} + \vec{E} \bullet \vec{\sigma},
\label{urb-E-Pauli}
\end{equation}
where
\begin{eqnarray}
\vec{E} \bullet \vec{\sigma} &=& E_{x} {\sigma}_{x} +
E_{y}{\sigma}_{y} + E_{z}{\sigma}_{z}, \label{urb-E-s}\\
E_{0} &=& \frac{1}{2}(E_{11} + E_{22}), \label{urb-E-0}\\
E_{z} &=& \frac{1}{2}(E_{11} - E_{22}), \label{urb-E-z}\\
E^{2} &=& \vec{E} \bullet \vec{E}
= E_{x}^{2} + E_{y}^{2} + E_{z}^{2} \label{urb-E} \\
& \equiv & E_{12}\,E_{21} + E_{z}^{2},\nonumber
\end{eqnarray}
where ${\sigma}_{\nu}, \; (\nu = x,y,z),$ are the Pauli matrices.

Solutions  of  Eq. (\ref{urb-E-eigen}), i.e., the eigenvalues
$\zeta$ for ${\bf E}$ can be expressed in terms of matrix elements
$E_{jk}$ of ${\bf E}$:
\begin{eqnarray}
{\zeta}_{\pm} & = & \frac{1}{2} \Big( E_{11} + E_{22} \pm \{
[E_{11} - E_{22}]^{2} \; + \; 4E_{12} E_{21} {\}}^{1/2}
\Big) \label{urb-ksi-pm} \\
& \equiv & E_{0} \pm E , \label{urb-E-pm-E}
\end{eqnarray}
The eigenstates of ${\bf E}$ are linear combinations of
time--independent  vectors  $|{\bf  1}\rangle  $  and  $|{\bf
2}\rangle  $:
\begin{equation}
|e_{\pm}\rangle   = p_{\pm}|{\bf 1}\rangle   + q_{\pm}|{\bf
2}\rangle   . \label{urb-e-pm}
\end{equation}
The coefficients of such  combinations  are  determined  by  the
solution to the matrix equation:
\begin{equation}
[{\bf E} - {\zeta}_{\pm} I_{\parallel} ] \left( \begin{array}{c}
p_{\pm} \\ q_{\pm}
\end{array} \right)
= 0 , \label{urb-Eq-E}
\end{equation}
from which one obtains
\begin{equation}
\frac{q_{\pm}}{p_{\pm}} = - \frac{E_{11} - {\zeta}_{\pm}}{E_{12}}
\equiv - \frac{E_{21}}{E_{22} - {\zeta}_{\pm}}. \label{urb-p/q}
\end{equation}
There are two, in general, nonorthogonal eigenvectors for ${\bf
E}$:
\begin{equation}
|e_{\pm}\rangle   \equiv p_{\pm} (\; |{\bf 1}\rangle   -
{\alpha}_{\pm}|{\bf 2}\rangle   ) , \label{urb-e-pm-1}
\end{equation}
where
\begin{eqnarray}
{\alpha}_{\pm}  = - \frac{q_{\pm}}{p_{\pm}} &\equiv&
\frac{E_{11} - {\zeta}_{\pm}}{E_{12}}  \label{urb-alpha-pm}  \\
& \equiv & \frac{E_{z} \mp E}{E_{12}} . \nonumber
\end{eqnarray}
Requiring that $|e_{\pm}\rangle  $ be normalized, one obtains
\begin{displaymath}
|p_{\pm}|^{2} = \frac{ 1}{1 + |{\alpha}_{\pm}|^{2}} .
\end{displaymath}
Eigenvectors and eigenvalues for $H_{\parallel}$ and
$U_{\parallel}$ can  be  obtained inserting into these formulae
matrix elements of $H_{\parallel}$ and  $U_{\parallel}$ instead of
${\bf E}$.

Note that in the general case neither eigenvectors for $H_{||}$
nor for $U_{||}$ are orthogonal due to the nonhermicity of
$H_{||}$ and nonunitarity of $U_{||}$.

\subsection{Definitions and properties of {\cal C}, {\cal  P}
and {\cal T} \\ transformations.}

Now let us discuss some consequences of  the  assumed  CP-- or
CPT--invariance  of  the  system   under   considerations, i.e.,
implications of the following assumptions
\begin{equation}
[{\cal C}{\cal P}, H] = [{\cal C}{\cal P}, P] = 0,
\label{urb-[CP,H]}
\end{equation}
and
\begin{equation}
[ \Theta , H ]  =  0, \label{urb-[CPT,H]}
\end{equation}
\begin{equation}
[ \Theta , P ]  =  0, \label{urb-[CPT,P]}
\end{equation}
where ${\Theta} = {\cal C}{\cal P}{\cal  T}$  and  ${\cal  C},  \;
{\cal P}$ and ${\cal T}$ denote operators realizing charge
conjugation, parity and  time  reversal  respectively, for vectors
in  $\cal H$.

Under transformations: unitary $\cal P$, $\cal C$, ${\cal C}{\cal
P}$, and antiunitary: $\cal T$, $\Theta$, the vectors, say
$|{\Psi}_{k};{\bf p_{k}}, {\bf \lambda}_{k} \rangle  \in {\cal
H}$, where ${\bf p}_{k}$ and  ${\bf  \lambda}_{k}$  denote
momentum and spin  of particles ${\Psi}_{k}$ respectively, behave
as follows \cite{ryder,werle}

\begin{eqnarray}
{\cal C}| {\Psi}_{k}; {\bf p}_{k}, {\bf \lambda}_{k} \rangle   & =
& {\beta}_{k}^{C}|{\overline {\Psi}_{k}};
{\bf p}_{k}, {\bf \lambda}_{k} \rangle   , \nonumber \\
{\cal P}| {\Psi}_{k}; {\bf p}_{k}, {\bf \lambda}_{k} \rangle   & =
& {\beta}_{k}^{P}|{\Psi}_{k};
- {\bf p}_{k}, - {\bf \lambda}_{k} \rangle   , \nonumber \\
{\cal C}{\cal P}| {\Psi}_{k}; {\bf p}_{k}, {\bf \lambda}_{k}
\rangle   & = & {\beta}_{k}^{CP}|{\overline {\Psi}_{k}};
- {\bf p}_{k}, - {\bf \lambda}_{k} \rangle   ,  \label{urb-CP-psi} \\
\langle {\cal T} {\Psi}_{j}|{\cal T} {\Psi}_{k}\rangle   & = &
\langle {\Psi}_{k}|{\Psi}_{j}\rangle   , \nonumber \\
\langle {\Theta} {\Psi}_{j}|{\Theta} {\Psi}_{k}\rangle   & \equiv
& \langle {\overline {\Psi}_{j}}|{\overline {\Psi}_{k}}\rangle   =
\langle {\Psi}_{k}|{\Psi}_{j}\rangle   , \label{urb-CPT-psi}
\end{eqnarray}
where: $|{\beta}_{k}^{C}|  =  |{\beta}_{k}^{CP} | =  1$ and
${\overline {\Psi}_{k}}$ denotes an antiparticle for ${\Psi}_{k}$.
There is ${\cal C}{\cal P} = ({\cal C}{\cal P})^{+} = ({\cal C}
{\cal P})^{-1}$ and therefore eigenvalues for ${\cal C}{\cal P}$
are equal $\pm  1$. The parity of two-particle state
$|{\Psi}_{1},{\Psi}_{2}; {\bf p}_{1}, {\bf \lambda}_{1}; {\bf
p}_{2},{\bf p}_{2};L \rangle  $, where $L$ is a relative angular
momentum, equals
\begin{equation}
{\cal P}|{\Psi}_{1},{\Psi}_{2}; {\bf p}_{1}, {\bf \lambda}_{1};
{\bf p}_{2},{\bf p}_{2};L \rangle   =
{\beta}_{1}^{P}{\beta}_{2}^{P} ( - 1 )^{L} |{\Psi}_{1},{\Psi}_{2};
- {\bf p}_{1}, - {\bf \lambda}_{1}; - {\bf p}_{2}, - {\bf
\lambda}_{2};L \rangle  . \label{urb-P-psi}
\end{equation}
The parity of one pion state is  ${\beta}_{\pi}^{P}  =  -  1$  and
therefore   properties   (\ref{urb-CP-psi}), (\ref{urb-P-psi})
imply that there    is ${\beta}_{\pi}^{P}
{\beta}_{\pi}^{P}(-1)^{L}$ $\stackrel{\rm def}{=} {\beta}_{\pi ,
\pi}^{P}$ $\equiv  + 1 $ for the two-pion state with a zero
relative angular momentum $L$. This means that \cite{ryder}
\begin{equation}
{\cal C}{\cal P}|\pi , \pi ; L = 0\rangle   = ( + 1) |\pi , \pi ;
L = 0\rangle   . \label{urb-CP-2pi}
\end{equation}
Let $L$ be the relative momentum of a two-pion  subsystem  and
let $L'$ be the angular momentum of third pion about the center of
mass of the two-pion subsystem then \cite{ryder}
\begin{equation}
{\cal C}{\cal P}|\pi , \pi ,\pi ; L = 0, L' = 0\rangle   = ( -
1)^{1 + L'= 0} |\pi , \pi ,\pi ; L = 0, L' = 0\rangle   .
\label{urb-CP-3pi}
\end{equation}
The following phase  convention  for  neutral  kaons is commonly
used
\begin{equation}
{\cal C}{\cal P}|{\bf 1}\rangle   = ( - 1)|{\bf 2}\rangle  ,
\;\;\;\;\; \; {\cal C}{\cal P}|{\bf 2}\rangle   = ( - 1) |{\bf
1}\rangle  , \label{urb-CP-K}
\end{equation}
and
\begin{equation}
\Theta |{\bf 1}\rangle   = e^{\textstyle{-i\theta}}|{\bf 2}\rangle
, \;\;\;\;\; \; \Theta|{\bf 2}\rangle   = e^{\textstyle{-i\theta}}
|{\bf 1}\rangle  . \label{urb-CPT1=2}
\end{equation}
Note that vectors
\begin{equation}
|K_{1(2)}\rangle \stackrel{\rm def}{=} \frac{1}{\sqrt{2}}(\;|{\bf
1}\rangle - (+) |{\bf 2}\rangle ) , \label{urb-K12}
\end{equation}
are normalized, orthogonal
\begin{equation}
\langle K_{j}|K_{k}\rangle  = {\delta}_{jk}, \; \; \; (j,k =1,2) ,
\label{urb-K-12-a}
\end{equation}
eigenvectors of ${\cal C}{\cal P}$ transformation,
\begin{equation}
{\cal CP} |K_{1(2)}\rangle = + (- 1)|K_{1(2)}\rangle,
\label{urb-CP+-1}
\end{equation}
for the eigenvalues $ + 1$ and $-1$ respectively.

Now let us analyze some general properties of the neutral kaon
complex caused by CP symmetry and independent of the approximation
used to describe time evolution in neutral kaon complex.
%LOY theory.
The conservation of %${\cal C}{\cal  P}$
CP implies  that the  decay modes, and hence the lifetimes of
$K_{1}$  and $K_{2}$  must  be different. The mass of $K_{0}$ is
$m_{K} \simeq  498$  MeV,  the masses  of  pions  are
$m_{{\pi}^{\pm}} \simeq  140$   MeV   and $m_{{\pi}^{0}} \simeq
135 $ MeV, so two and three pions decays are energetically allowed
for neutral kaons.  If  ${\cal  C}{\cal P}$ symmetry  is
conserved  then only  allowed  decay   are $K_{1} \;
\longrightarrow \; 2\pi$, $K_{2} \; \longrightarrow  \; 3\pi$.
$K_{2}$ cannot decay in two pions at all. These conclusions can be
easily  derived  from properties   of   decay   amplitudes $A(K
\rightarrow 2\pi ), \; A({\overline K} \rightarrow 2 \pi )$, etc..
Defining
\begin{equation}
A(K \rightarrow 2\pi ) \stackrel{\rm def}{=} \langle \pi \pi
|H|{\bf 1}\rangle ,\;\; \; \; A({\overline K} \rightarrow 2 \pi )
\stackrel{\rm def}{=} \langle  \pi \pi |H|{\bf 2}\rangle    ,
\label{urb-A-K-2pi}
\end{equation}
and so on, and using relations (\ref{urb-CP-2pi}),
(\ref{urb-CP-K}) one finds in  the case  of conserved CP--symmetry
\begin{eqnarray}
A(K \rightarrow 2\pi ) &=& \langle \pi \pi |H|{\bf 1}\rangle
\label{urb-A-K-2pi-a} \\
& \equiv & \langle  \pi \pi |({\cal C} {\cal P} )^{-1} ({\cal C}
{\cal P}) H ({\cal C} {\cal P})^{-1} ({\cal C} {\cal P}) |{\bf
1}\rangle
\nonumber \\
&=& \langle  {\cal C}{\cal P} ( \pi \pi )|{[} ({\cal C} {\cal P})
({\cal C} {\cal P})^{-1} {]} H ( {\cal C} {\cal P} )
|{\bf 1}\rangle    )  \nonumber  \\
&=& - \langle  \pi \pi |H|{\bf 2}\rangle     \equiv - A({\overline
K} \rightarrow 2 \pi ). \label{urb-A-K-2pi-b}
\end{eqnarray}
Analogous considerations lead to the property
\begin{equation}
A(K \rightarrow 3\pi ) = A({\overline K} \rightarrow 3\pi ).
\label{urb-A-K-3pi}
\end{equation}
Using these relations, the decay amplitudes $K_{1(2)} \;
\longrightarrow 2 \pi$ can be derived as
\begin{eqnarray}
A(K_{1} \rightarrow 2 \pi ) & \equiv & \langle  \pi \pi
|H|K_{1}\rangle
\label{urb-A-K1-2pi} \\
& \equiv & 2^{-1/2}[A(K  \rightarrow  2  \pi)  -  A({\overline  K}
\rightarrow 2 \pi )] \nonumber \\
&=& 2^{1/2} A(K \rightarrow 2 \pi ), \label{urb-A-K1-2pi-a}
\end{eqnarray}
\begin{eqnarray}
A(K_{2} \rightarrow 2 \pi ) & \equiv & \langle  \pi \pi
|H|K_{1}\rangle
\label{urb-A-K2-2pi} \\
& \equiv & 2^{-1/2} [ A(K  \rightarrow  2  \pi)  +  A({\overline
K} \rightarrow 2 \pi ) ]  \equiv 0. \label{urb-A-K2-2pi-a}
\end{eqnarray}
For three pion decays mode one obtains in a similar manner
\begin{eqnarray}
A(K_{1} \rightarrow 3 \pi ) & \equiv & \langle  \pi  \pi \pi
|H|K_{1}\rangle
\equiv 0,  \label{urb-A-K1-3pi} \\
A(K_{2} \rightarrow 3 \pi ) & \neq & 0. \nonumber
\end{eqnarray}
Thus we see that certain decay modes which are allowed for $K_{1}$
are forbidden for $K_{2}$ and vice versa. This is the essence of
Gell-- Mann  ad  Pais  theory  \cite{ryder,werle,gell-mann}.  The
difference in the allowed transitions implies a corresponding
difference  in the lifetimes $\tau$. It should be ${\tau}_{K_{1}}
\ll {\tau}_{K_{2}}$ because ${\tau}_{K_{2}} \sim (m_{K} - 3
m_{\pi})^{-1}$  and ${\tau}_{K_{1}} \sim (m_{K}  -  2
m_{\pi})^{-1}$, i.e.,  the decay rate for two pion decay is larger
then for three pion decays.

\subsection{Properties of eigenvectors for $H_{||}$.}

Generally, in the case of two dimensional subspace ${\cal H}_{||}$
the eigenvectors of $H_{\parallel}$ acting in this ${\cal H}_{||}$
will be denoted as $|l\rangle , |s\rangle  $. The solutions of the
eigenvalue problem (\ref{urb-E-eigen}) can be easily adopted to
this case by identifying eigenvectors  $|e_{\pm}\rangle $ of ${\bf
E}$ with $ |l(s)\rangle $,
\begin{eqnarray}
|e_{+(-)}\rangle  \; \rightarrow  |l(s)\rangle   & = &
\frac{1}{\sqrt{|p_{l(s)}|^{2} + |q_{l(s)}|^{2} }}\;
\Big( p_{l(s)} |{\bf 1} \rangle  + q_{l(s)} |{\bf 2} \rangle  \Big)
\label{urb-ls} \\
& \equiv & {\rho}_{l(s)} \Big(|{\bf 1}\rangle  -
{\alpha}_{l(s)}|{\bf 2}\rangle  \Big) , \label{urb-ls-12}
\end{eqnarray}

\begin{equation}
\frac{q_{l(s)} }{ p_{l(s)} } = - \frac{ h_{11} - {\mu}_{l(s)} }{
h_{12} } \equiv - \frac{ h_{21} }{ h_{22} - {\mu}_{l(s)} },
\label{urb-q/p-ls}
\end{equation}

\begin{equation}
{\alpha}_{l(s)} = \frac{h_{z} - (+) h}{h_{12}},
\label{urb-alpha-ls}
\end{equation}
{\noindent}and replacing eigenvalues $\zeta_{\pm}$ of ${\bf E}$ by
${\mu}_{l(s)}$, i.e. by eigenvalues  of $H_{||}$ for eigenvectors
$|l(s)\rangle $,
\begin{equation}
{\zeta}_{+(-)} \; \rightarrow \; {\mu}_{l(s)} = h_{0} +(-) h
\equiv  m_{l(s)}  - \frac{i}{2}  {\gamma}_{l(s)},
\label{urb-mu-ls}
\end{equation}
where $m_{l(s)}, {\gamma}_{l(s)}$ are real, and
\begin{eqnarray} h_{0} & = & \frac{1}{2}(h_{11} + h_{22}),
\label{urb-h-0} \\ h &
\equiv  &  \sqrt{ h_{z}^{2} + h_{12} h_{21} }, \label{urb-h-a} \\
h_{z} & = &  \frac{1}{2} (h_{11}  -  h_{22}).   \label{urb-h-z}
\end{eqnarray} In the case of neutral kaons these
eigenvectors correspond to the long (vector $|l\rangle $) and
short (vector $|s\rangle $) living superpositions of $K_{0}$
and $\overline{K_{0}}$.

The  following identity taking place for ${\mu}_{l}$ and
${\mu}_{s}$ will be needed in next Sections:
\begin{eqnarray}
{\mu}_{l}  +  {\mu}_{s}  &  =  & h_{11} + h_{22} ,
\label{urb-mu-l+mu-s}
\\ {\mu}_{s} - {\mu}_{l} &  =  &  2h \stackrel{\rm def}{=}  \Delta
\mu,  \label{new6}  \\  {\mu}_{l}  \, {\mu}_{s}  &  =  &  h_{11}
h_{22}  -   h_{12}h_{21} \equiv \det\, H_{||} ,
\label{urb-mu-l-mu-s}
\end{eqnarray}

Using the  eigenvectors $|K_{1(2)}\rangle $, (\ref{urb-K12}), of
the CP--transformation  for the eigenvalues $\pm  1$, vectors
$|l\rangle $ and $|s\rangle $ can be expressed  as follows
\cite{cronin,comins,dafne,barmin}
\begin{equation} |l(s)\rangle  \equiv \frac{1}{\sqrt{1  +
|{\varepsilon}_{l(s)}|^{2} }} \;\Big(\;|K_{2(1)} \rangle   +
{\varepsilon}_{l(s)} |K_{1(2)} \rangle \Big) , \label{urb-ls-K12}
\end{equation}
where
\begin{eqnarray}  {\varepsilon}_{l}  &  =   &
\frac{h_{12} - h_{11} +  {\mu}_{l}}{h_{12}  +  h_{11}  -
{\mu}_{l}} \equiv - \frac{h_{21}  -  h_{22}  +  {\mu}_{l}}{h_{21}
+  h_{22}  - {\mu}_{l}}, \label{urb-eps-l} \\ {\varepsilon}_{s} &
= & \frac{h_{12}  + h_{11}  -  {\mu}_{s}}{h_{12}  -  h_{11}  +
{\mu}_{l}}   \equiv   - \frac{h_{21} + h_{22} - {\mu}_{s}}{h_{21}
-  h_{22}  +  {\mu}_{s}}, \label{urb-eps-s} \end{eqnarray} This
form of $|l\rangle  $ and $|s\rangle  $ is used in many papers
when possible departures from CP-- or  CPT--symmetry  in the
system considered are discussed. The  following  parameters  are
used to describe the scale  of CP-- and possible  CPT  --
violation effects \cite{cronin,comins,dafne,barmin}:
\begin{equation}
\varepsilon \stackrel{\rm def}{=} \frac{1}{2} (  {\varepsilon}_{s}
+ {\varepsilon}_{l}  ) , \label{urb-eps}  \end{equation}
\begin{equation}  \delta \stackrel{\rm def}{=} \frac{1}{2} (
{\varepsilon}_{s}  - {\varepsilon}_{l}  )  . \label{urb-delta}
\end{equation} According  to the   standard interpretation,
$\varepsilon$ describes violations of CP--symmetry and  $\delta$
is considered as  a  CPT--violating parameter
\cite{LOY2,cronin,comins,dafne,barmin}.  Such an interpretation of
these parameters  follows from properties of LOY theory of  time
evolution  in  the subspace   of neutral kaons {\cite{LOY1} ---
\cite{Bigi}. We have
\begin{eqnarray} \varepsilon & = & \frac{h_{12}  -
h_{21}}{D} \label{urb-eps-D} \\ \delta & = & \frac{h_{11}  -
h_{22} }{D} \equiv  \frac{2  h_{z}  }{D}  ,  \label{urb-delta-D}
\end{eqnarray} where
\begin{equation} D  \stackrel{\rm def}{=}  h_{12} + h_{21} +  \Delta
\mu . \label{urb-D} \end{equation}

Starting from  Eqs.  (\ref{urb-mu-l+mu-s})
--- (\ref{urb-mu-l-mu-s}) and  (\ref{urb-eps-l}),
(\ref{urb-eps-s})  and  using  some
known identities for ${\mu}_{l}, {\mu}_{s}$  one can express
matrix elements $h_{jk}$  of  $H_{\parallel}$  in  terms  of  the
physical parameters      ${\varepsilon}_{l(s)}$      and
${\mu}_{l(s)}$:
\begin{eqnarray} h_{11} & =  &  \frac{{\mu}_{s}  +  {\mu}_{l}}{2}  +
\frac{{\mu}_{s}   -    {\mu}_{l}}{2}    \frac{{\varepsilon}_{s} -
{\varepsilon}_{l}}{ 1  -  {\varepsilon}_{l}  {\varepsilon}_{s} }
, \label{urb-h11-mu} \\ h_{22} & =  &  \frac{{\mu}_{s}  +
{\mu}_{l}}{2}  - \frac{{\mu}_{s}   -    {\mu}_{l}}{2}
\frac{{\varepsilon}_{s}    - {\varepsilon}_{l}}{ 1  -
{\varepsilon}_{l}  {\varepsilon}_{s}  }  , \label{urb-h22-mu} \\
h_{12}  &  =  &  \frac{{\mu}_{s}  -  {\mu}_{l}}{2} \frac{(1  +
{\varepsilon}_{l})(1  +  {\varepsilon}_{s})}  {   1   -
{\varepsilon}_{l}{\varepsilon}_{s} } , \label{urb-h12-mu} \\
h_{21} & = & \frac{{\mu}_{s} - {\mu}_{l}}{2} \frac{(1  -
{\varepsilon}_{l})(1  - {\varepsilon}_{s})} { 1  -
{\varepsilon}_{l}{\varepsilon}_{s}  }  . \label{urb-h21-mu}
\end{eqnarray} These relations lead to  the  following equations
\begin{eqnarray}  h_{11}  -  h_{22}  &  =  &  \Delta  \mu
\frac{{\varepsilon}_{s} - {\varepsilon}_{l}}{ 1 -
{\varepsilon}_{l} {\varepsilon}_{s} } , \label{urb-h11-h22-mu} \\
h_{12} + h_{21} & = &  \Delta \mu   \frac{1   +{\varepsilon}_{l}
{\varepsilon}_{s}   }{   1    - {\varepsilon}_{l}
{\varepsilon}_{s} } , \label{urb-h12+h21-mu}  \\  h_{12}  - h_{21}
&   = &    \Delta    \mu    \frac{{\varepsilon}_{s}    +
{\varepsilon}_{l}}{ 1  -  {\varepsilon}_{l}  {\varepsilon}_{s}  }
, \label{urb-h12-h21-mu} \end{eqnarray} Note that relations
(\ref{urb-h11-mu})
--- (\ref{urb-h12-h21-mu}) are valid for arbitrary values
of ${\varepsilon}_{l(s)}$.  From  (\ref{urb-h11-h22-mu})  one
infers  that if $\Delta \mu \neq  0$  then:  \begin{equation}
h_{11}  = h_{22}  \; \Longleftrightarrow   \;{\varepsilon}_{l} =
{\varepsilon}_{s}. \label{urb-h11=h22-eps} \end{equation} Relation
(\ref{urb-h12-h21-mu})  enables  us  to conclude     that
parameters ${\varepsilon}_{l}$      and ${\varepsilon}_{s}$ need
not be small, in order  that  $\varepsilon  =  0$ (\ref{urb-eps-D}).
Indeed, the identity  (\ref{urb-h12-h21-mu})  implies  that  for
$\Delta \mu  \neq 0$
\begin{equation}   h_{12}   =   h_{21}   \; \Longleftrightarrow
\;  {\varepsilon}_{l}  =  -  {\varepsilon}_{s},
\label{urb-h12=h21-eps}
\end{equation} for any values of $|{\varepsilon}_{l}|,
|{\varepsilon}_{s}|$.

It is appropriate to emphasize at  this  point that all relations
(\ref{urb-h11-mu}) --- (\ref{urb-h12=h21-eps}) do not depend  on a
special form of the effective  Hamiltonian  $H_{\parallel}$.  They
are induced by geometric relations between various base  vectors
in two--dimensional subspace ${\cal H}_{\parallel}$. On the other
hand, the interpretation of the above relations depends on the
properties of  the matrix   elements   $h_{jk}$   of the effective
Hamiltonian $H_{\parallel}$, i.e., if for example $H_{\parallel}
\neq  H_{LOY}$, where  $H_{LOY}$  is  the  LOY effective
Hamiltonian,   then   the interpretation of $\varepsilon$
(\ref{urb-eps}) and  $\delta$  (\ref{urb-delta}) etc., need not be
the same for $H_{\parallel}$  and  for $H_{LOY}$.

Experimentally measured  values  of  parameters
${\varepsilon}_{l}, {\varepsilon}_{s}$  are  very  small  for
neutral  kaons.  Assuming
\begin{equation}    |{\varepsilon}_{l}|     \ll     1,     \;     \;
|{\varepsilon}_{s}|  \ll  1,   \label{urb-eps<<1}   \end{equation}
from ({\ref{urb-h11-h22-mu}) one finds: \begin{equation} h_{11}  -
h_{22} \simeq ({\mu}_{s} - {\mu}_{l}) ({\varepsilon}_{s}  -
{\varepsilon}_{l}  ), \label{urb-h11-h22-mu-eps}    \end{equation}
and ({\ref{urb-h12+h21-mu})     implies
\begin{equation} h_{12}  +  h_{21}  \simeq  {\mu}_{s}  -  {\mu}_{l},
\label{urb-h12+h21-mu-eps}     \end{equation}     and
(\ref{urb-h12-h21-mu}) gives
\begin{equation} h_{12} -  h_{21}  \simeq  ({\mu}_{s}  -  {\mu}_{l})
({\varepsilon}_{s}     +      {\varepsilon}_{l}).
\label{urb-h12-h21-mu-eps}
\end{equation} Relation (\ref{urb-h12+h21-mu-eps}) means that in  the  considered
case   of   small   values   of   parameters $|{\varepsilon}_{l}|,
|{\varepsilon}_{s}|$ (\ref{urb-eps<<1}), the quantity  $D$
(\ref{urb-D}) appearing in the formulae for $\delta$ and
$\varepsilon$  approximately equals \begin{equation} D \simeq
2({\mu}_{s} - {\mu}_{l})  \equiv  2 \Delta \mu .
\label{urb-D-mu-eps}
\end{equation}

Keeping  in  mind  that $h_{jk}  =  M_{jk}  -   \frac{i}{2}
{\Gamma}_{jk}   ,   M_{kj}   = M_{jk}^{\ast},{\Gamma}_{kj}   =
{\Gamma}_{jk}^{\ast}$   and   then starting form Eqs.
(\ref{urb-h11-h22-mu-eps}) ---  (\ref{urb-h12-h21-mu-eps})  and
separating real and imaginary  parts  one  can  find  some  useful
relations:
\begin{eqnarray} 2{\Re \,} (M_{12})  &  \simeq  &  m_{s}  -  m_{l},
\label{urb-Re-M12} \\ 2{\Re \,}({\Gamma}_{12}) & \simeq &
{\gamma}_{s} - {\gamma}_{l}, \label{urb-Re-G12}  \\  2{\Im
\,}(M_{12})  & \simeq & - ({\gamma}_{s} - {\gamma}_{l})\, \Big[\,
{\Im \,} (\frac{{\varepsilon}_{s}  + {\varepsilon}_{l}}{2})      +
\tan \, {\phi}_{SW}\; {\Re \,}(\frac{{\varepsilon}_{s}   +
{\varepsilon}_{l}}{2}   )\,\Big]    , \label{urb-Im-M12} \\
{\Im \,}({\Gamma}_{12}) & \simeq & -  ({\gamma}_{s} -
{\gamma}_{l})\, \Big[\, \tan \, {\phi}_{SW}\; {\Re \,}
(\frac{{\varepsilon}_{s} + {\varepsilon}_{l}}{2})   -  {\rm  Im}
(\frac{{\varepsilon}_{s}  + {\varepsilon}_{l}}{2})\,\Big] ,
\label{urb-Im-G12} \end{eqnarray} etc., where $\Re \, (z)$ and
$\Im \, (z)$ denote the real and imaginary parts of $z$
respectively, and
\begin{equation}    \tan \, {\phi}_{SW}    \stackrel{\rm     def}{=}
\frac{2(m_{l} - m_{s})}{{\gamma}_{s} - {\gamma}_{l}} .
\label{urb-tan-phi}
\end{equation} and \begin{eqnarray}  {\Re \,}(h_{11}  -  h_{22})  &
\equiv & M_{11} - M_{22} = M_{1} - M_{2} \nonumber \\ & \simeq & -
({\gamma}_{s} - {\gamma}_{l}) \Big[\, \tan \,{\phi}_{SW}\; {\Re
\,} (\frac{{\varepsilon}_{s} - {\varepsilon}_{l})}{2})   \nonumber
\\   &    \;    &    -    {\Im \,}(\frac{{\varepsilon}_{s}     -
{\varepsilon}_{l})}{2}) \,\Big], \label{urb-Re-h11-h22} \\ - {\Im
\,}(h_{11} - h_{22} ) & \equiv & \frac{1}{2} ({\Gamma}_{11}  -
{\Gamma}_{22} ) \nonumber   \\   &   \simeq   & ({\gamma}_{s} -
{\gamma}_{l})\, \Big[\, {\Re \,} (\frac{{\varepsilon}_{s}  -
{\varepsilon}_{l})}{2}) \nonumber \\ & \; & + \tan \,{\phi}_{SW}
\; {\Im \,}(\frac{{\varepsilon}_{s}     - {\varepsilon}_{l})}{2})
\,\Big], \label{urb-Im-h11-h22}  \end{eqnarray}  etc.. One should
remember that relations  (\ref{urb-Re-M12})  ---
(\ref{urb-Im-G12})   and (\ref{urb-Re-h11-h22}),
(\ref{urb-Im-h11-h22}) are valid only  if condition
(\ref{urb-eps<<1}) holds. Completing the system of these last six
relations  one  can rewrite Eq. (\ref{urb-mu-l+mu-s}) to obtain
\begin{eqnarray} M_{11} + M_{22}  &  =  & m_{l} + m_{s},
\label{urb-M11+M22} \\ {\Gamma}_{11} + {\Gamma}_{22} & =  &
{\gamma}_{l} + {\gamma}_{s}  .  \label{urb-G11+G22}
\end{eqnarray} These last two Equations are exact independently of
whether the condition (\ref{urb-eps<<1})  holds  or  not.

\section{Lee, Oehme and Yang model.}

\subsection{Lee, Oehme and Yang approximation.}

The source of The Lee, Oehme and Yang (LOY) approximation for
decay  of neutral kaons is the well known Weisskopf--Wigner
approach  to  a description  of  unstable  states.  Within  this
approach,   the Hamiltonian $H$ for the problem is divided into
two parts $H^{(0)}$ and $H^{(1)}$ such that $|K_{0}\rangle  \equiv
|{\bf 1}\rangle$ and $|{\overline K}_{0}\rangle \equiv |{\bf
2}\rangle $ are twofold degenerate eigenstates of $H^{(0)}$ to the
eigenvalue $m_{0}$,
\begin{equation}
H^{(0)} |{\bf j} \rangle = m_{0} |{\bf j }\rangle, \; \;  j = 1,2
; \label{urb-Hm-0}
\end{equation}
and $H^{(1)} \equiv H - H^{(0)}$ induces transitions from these
states to other (unbound)  eigenstates $|\varepsilon \rangle $ of
$H^{(0)}$ and consequently also between $|K_{0}\rangle $ and $|
{\overline K}_{0} \rangle $. So, the problem which one usually
considers is the time evolution of a state which is prepared
initially as a superposition of $|K_{0}\rangle $ and $| {\overline
K}_{0}\rangle $  states.

In the kaon rest--frame, this time evolution  is  governed  by the
Schr\"{o}dinger equation (\ref{urb-Schrod}). Solutions, $|\psi ; t
\rangle $, (\ref{urb-psi-gen}), of this Equation have the form
\[  |{\psi};t\rangle   = a_{1}(t)|{\bf 1}\rangle
+ a_{2}(t)|{\bf 2}\rangle   + \sum_{j} F_{j}(t) |F_{j}\rangle  ,
\] where $|F_{j}\rangle  $ $\equiv$ $\sum_{\varepsilon}  \langle
\varepsilon |F_{j}\rangle  | \varepsilon \rangle  $ $=$
$\sum_{\varepsilon} f_{j}( \varepsilon )| \varepsilon \rangle \in
{\cal H}$ represents the decay products in the channel $j$;
$\langle \varepsilon|{\bf k}\rangle $ $=$ $0$, $k = 1,2$. It is
assumed that $F_{j}(0) = 0$.

Using the interaction representation  and  rescaling  respectively
energy $\varepsilon$: defining  $\omega  =  \varepsilon  -  m_{0}$
which means that the zero of energy is taken to be rest energy  of
$K$, instead of the Schr\"{o}inger equation (2) for $|\psi
;t\rangle  $, Lee,  Oehme  and  Yang  obtained  the  following
equations   for amplitudes $a_{1}(t)$, $a_{2}(t)$ and
$F_{j}(\omega ,t)$ replacing $F_{j}(t)$ \cite{LOY1},
\cite{Gaillard}, (for details see \cite{Urb-Pisk-2000}),
\begin{equation}
i \frac{\partial}{\partial t} a_{k}(t) = \sum_{l = 1}^{2}
H_{kl}a_{l}(t) + \sum_{j, \omega} H_{kj}( \omega ) F_{j}( \omega
,t) e^{\textstyle -i \omega t}, \; \; \; \; (k = 1,2)
\label{urb-Eq-a(t)}
\end{equation}
\begin{equation}
i \frac{\partial}{\partial t} F_{j}( \omega ,t) = e^{\textstyle i
\omega t} [ H_{j1}( \omega ) a_{1}(t) +  H_{j2}( \omega ) a_{2}(t)
], \label{urb-Eq-F(t)}
\end{equation}
where $H_{kj}( \omega ) =  H_{jk}( \omega )^{\ast}$, ($k =1,2$),
are the matrix elements responsible for the decay  and
\begin{equation}
H_{kl}= \langle {\bf k}|H|{\bf l}\rangle \;\;\;\;(k,l =  1,2).
\label{urb-H-jk}
\end{equation}
Equations (\ref{urb-Eq-a(t)}) are exact, Equation
(\ref{urb-Eq-F(t)}) has been obtained by ignoring  the series
containing matrix elements of type $\langle \varepsilon |H^{(1)}|
{\varepsilon}'\rangle  $. (Eq. (\ref{urb-Eq-F(t)}) is exact for
the  model  in which $\langle \varepsilon
|H^{(1)}|{\varepsilon}'\rangle  $  $=$ $0$
---  if these matrix elements are very small, then this equation
is  be treated as a very good approximation for the model
considered).

The boundary conditions for Eqs. (\ref{urb-Eq-a(t)}),
(\ref{urb-Eq-F(t)}) are the following:
\begin{equation}
a_{k}(0) \neq 0, \;\;\;\;(k = 1, 2), \label{a_k(0)}
\end{equation}
and
\begin{equation}
F_{j}( \omega ,0) = 0. \label{urb-F(0)}
\end{equation}

To  solve  the  system  of  coupled  Equations
(\ref{urb-Eq-a(t)}), (\ref{urb-Eq-F(t)}),   the exponential time
dependence for  amplitudes $a_{k}(t)$  has  been assumed in
\cite{LOY1}, i.e., it has been assumed that
\begin{equation}
\frac{ a_{1}(t) }{ a_{1}(0) } = \frac{ a_{2}(t) }{ a_{2}(0) }
\equiv e^{ {\textstyle - \frac{\Lambda t}{2} } }, \; \; \;\;{\Re
\,} (\Lambda) \, > \, 0. \label{urb-LOY-main}
\end{equation}

This crucial assumption is the essence of the approximation  which
was made in \cite{LOY1} and determines the properties  of  the
so--called LOY model of neutral kaons decay, i.e., the effective
Hamiltonian $H_{LOY}$ governing time  evolution  in  neutral
kaons  subspace, which is a consequence of Eqs.
(\ref{urb-Eq-a(t)}), (\ref{urb-Eq-F(t)}) and  of  the requirement
(\ref{urb-LOY-main}).

Inserting (\ref{urb-LOY-main}) into Eq. (\ref{urb-Eq-F(t)}) and
taking into account (\ref{urb-F(0)}), Eq. (\ref{urb-Eq-F(t)}) can
easily be solved to obtain
\begin{eqnarray}
iF_{j}( \omega ,t) & = & \Big[\, H_{j1}( \omega ) a_{1}(0) +
H_{j2}( \omega ) a_{2}(0) \,\Big] \frac{ e^{\textstyle i \omega t
- \Lambda t/2 } - 1 }{i  \omega  -
\Lambda /2 }  \nonumber  \\
& \equiv & -i \Big[\, H_{j1}( \omega ) a_{1}(t) +  H_{j2}( \omega
) a_{2}(t) \,\Big] e^{\textstyle i \omega t}\, \frac{1 -
e^{\textstyle \Lambda t/2} e^{\textstyle -i \omega t} } {  \omega
+ i {\Lambda \over 2}  }. \label{urb-F(omega,t)}
\end{eqnarray}
Next, one can eliminate the $F_{j}( \omega ,t)$ in
(\ref{urb-Eq-a(t)}) by substituting  (\ref{urb-F(omega,t)})  back
into (\ref{urb-Eq-a(t)}), and  then considering $\Lambda$ as a
very small number: $\Lambda \simeq 0$ one finds the following
equation for amplitudes $a_{k}(t)$  (for details,  the reader is
referred, e.g., to \cite{Gaillard,Urb-Pisk-2000})
\begin{equation}
i \frac{\partial}{\partial t} \left(
\begin{array}{c}
a_{1}(t) \\ a_{2}(t)
\end{array}
\right) = H_{LOY} \left(
\begin{array}{c}
a_{1}(t) \\ a_{2}(t)
\end{array}
\right) , \label{urb-LOY-Eq}
\end{equation}
where (see \cite{LOY1} --- \cite{Bigi}) $H_{LOY}$ $\equiv$
$M_{LOY} - {i \over 2} {\Gamma}_{LOY}$, and $M_{LOY} =
M_{LOY}^{+}$, $\Gamma_{LOY} = \Gamma_{LOY}^{+}$ are ($2 \times 2$)
matrices. Standard formulae for matrix elements $h_{kl}^{LOY}$
$\stackrel{\rm def}{=}$ $\langle {\bf k}|H_{LOY}|{\bf l}\rangle  $
$\equiv$ $M_{kl}^{LOY} - {i \over 2} {\Gamma}_{kl}^{LOY}$ can be
found, e.g., in \cite{LOY1}  --- \cite{Bigi}. One has
\begin{equation}
h_{jk}^{LOY} = H_{jk} - {\Sigma}_{jk}(m_{0}) \equiv M_{jk}^{LOY} -
\frac{i}{2}  {\Gamma}_{jk}^{LOY} , \; \; (j,k = 1,2 ),
\label{urb-h-jk-LOY}
\end{equation}
where ${\Sigma}_{jk}( \varepsilon )$ $=$ $\langle {\bf j}|\Sigma (
\varepsilon ) |{\bf k}\rangle $, and
\begin{eqnarray}
\Sigma ( \varepsilon ) & = & PHQ \frac{1}{ QHQ - \varepsilon - i0
}
QHP \label{urb-Sigma} \\
& \stackrel{\rm def}{=} & {\Sigma}^{R}( \varepsilon ) + i
{\Sigma}^{I}( \varepsilon ), \nonumber
\end{eqnarray}
and projectors $P,Q$ are defined formula (\ref{urb-P}). For
$\varepsilon$ real one finds ${\Sigma}^{R}( \varepsilon )$ $=$
${\Sigma}^{R}( \varepsilon )^{+}$ and ${\Sigma}^{I}( \varepsilon
)$ $=$ ${\Sigma}^{I}( \varepsilon )^{+}$ $\equiv$ ${1 \over 2}
\Gamma ( \varepsilon )$. Taking into account (\ref{urb-Hm-0}) one
can write that simply
\begin{eqnarray*}
M_{jk}^{LOY} & = & m_{0} {\delta}_{jk} + \langle {\bf
j}|H^{(1)}|{\bf k}\rangle    - \langle {\bf j}|HQ \; \; {\rm P.v.}
\frac{1}{Q H Q - m_{0}} \; \;
HQ|{\bf k} \rangle    ,  \\
{\Gamma}_{jk}^{LOY} & = &  2 {\pi}  \langle  {\bf j} |{H Q}  \;
{\delta} ({Q H Q} - m_{0} ) \;  {Q H}|{\bf k}\rangle    ,
\end{eqnarray*}
(here P.v. denotes a principal value). These formulae are the
frame for almost all calculations  of parameters characterizing
CP--violation effects in neutral  kaons decays, and  for searching
for possible  violations of CPT--symmetry, and for designing
CPT--violation tests in such a system \cite{LOY2}
--- \cite{Bigi}. The compact, operator form of $H_{LOY}$ is
\begin{equation}
H_{LOY} \equiv PHP - \Sigma (m_{0}) . \label{urb-H-LOY-op}
\end{equation}

$H_{LOY}$ acts in the subspace ${\cal H}_{\parallel} \stackrel{\rm
def}{=} P{\cal H}$ of ${\cal H}$ --- in the subspace of unstable
states $|{\bf 1}\rangle   , |{\bf  2}\rangle      \in {\cal
H}_{\parallel}$.  The  subspace  of  decay products ${\cal
H}_{\perp}$ is defined by the projector $Q$: ${\cal H}_{\perp}
\stackrel{\rm def}{=} Q{\cal H} \ni |F_{j}\rangle    $.

Using $H_{LOY}$ solutions of the evolution equation
(\ref{urb-LOY-Eq}) can  be written by means of an evolution
operator $U_{\parallel}^{LOY}(t)$ for subspace ${\cal
H}_{\parallel}$ as follows
\begin{equation}
\left(
\begin{array}{c}
a_{1}(t) \\ a_{2}(t)
\end{array}
\right) = U_{\parallel}^{LOY}(t) \left(
\begin{array}{c}
a_{1}(0) \\ a_{2}(0)
\end{array}
\right) , \label{urb-U-LOY-op}
\end{equation}
or
\begin{equation}
|\psi;t\rangle    _{\parallel} = U_{\parallel}^{LOY}(t) |\psi
\rangle    _{\parallel}, \label{urb-U-LOY-psi}
\end{equation}
where
\begin{eqnarray}
|\psi ;t\rangle_{\parallel} & \stackrel{\rm def}{=} & P |\psi
;t\rangle    \equiv a_{1}(t)|{\bf 1}\rangle     + a_{2}(t)|{\bf
2}\rangle, \label{urb-psi,t||}  \\
|\psi \rangle    _{\parallel} & = & |\psi ;t = 0 \rangle
_{\parallel} , \nonumber
\end{eqnarray}
and
\begin{eqnarray}
U_{\parallel}^{LOY} &=& \exp (-itH_{LOY}) \nonumber \\
& \equiv & e^{\textstyle -it\,h_{0}^{LOY}} \Big[ I_{\parallel}
\cos (t\,h^{LOY}) - i \frac{{\vec{h}}^{\;LOY} \bullet
\vec{\sigma}} {h^{LOY}} \sin (t\,h^{LOY}) \Big] .
\label{urb-U||-LOY-op}
\end{eqnarray}
Here the Pauli matrices representation is used (see
(\ref{urb-E-Pauli}) --- (\ref{urb-E})):
\begin{equation}
H_{LOY} \equiv h_{0}^{LOY} I_{\parallel} +  \vec{h}^{\;LOY}
\bullet \vec{\sigma}, \label{urb-H-LOY-oPauli}
\end{equation}
\[
h_{0}^{LOY}  =  \frac{1}{2} [ h_{11}^{LOY} + h_{22}^{LOY} ] ,
\]
\[
h_{z}^{LOY}  =  \frac{1}{2} [ h_{11}^{LOY} - h_{22}^{LOY} ] ,
\]
\begin{eqnarray*}
(h^{LOY} )^{2}  \equiv  \vec{h}^{\;LOY} \bullet \vec{h}^{\;LOY}
& = &  (h_{x}^{LOY} )^{2} + (h_{y}^{LOY} )^{2} + (h_{z}^{LOY} )^{2}   \\
&\equiv & h_{12}^{LOY} h_{21}^{LOY} + (h_{z}^{LOY} )^{2} .
\end{eqnarray*}

\subsection{Properties  of  neutral  kaons in the case  of \\
conserved CP--symmetry.}

Assuming that CP--symmetry is conserved in the system  considered,
i.e., that relation (\ref{urb-[CP,H]}) is valid, one finds that
\begin{equation}
[{\cal C}{\cal P}, PHP] = [{\cal C}{\cal P}, {\Sigma}(m_{0} ) ] =
0, \label{urb-[CP,PHP]}
\end{equation}
which means that
\begin{equation}
[{\cal C}{\cal P}, H_{LOY}] = 0. \label{urb-[CP,H-LOY]}
\end{equation}
Using this property and (\ref{urb-CP-K}), the matrix elements
$M_{jk}^{LOY}$ and ${\Gamma}_{jk}^{LOY}$ can be found as
\begin{equation}
M_{11}^{LOY} = M_{22}^{LOY} , \; \; {\Gamma}_{11}^{LOY} =
{\Gamma}_{22}^{LOY} , \label{urb-M11=M22-LOY}
\end{equation}
\begin{equation}
M_{12}^{LOY} = M_{21}^{LOY},  \;  \; {\Gamma}_{12}^{LOY} =
{\Gamma}_{21}^{LOY}, \label{urb-M12=M21-LOY}
\end{equation}
which give
\begin{equation}
h_{11}^{LOY} = h_{22}^{LOY} \equiv h_{0}^{LOY}, \; \;
h_{12}^{LOY} = h_{21}^{LOY}. \label{e41}
\end{equation}

These relations have the following consequences  for  eigenvectors
and  eigenvalues  of  $H_{LOY}$   in   the   case   of   conserved
CP--symmetry:
\begin{equation}
{\alpha}_{- (+)} \equiv {\alpha}_{2(1)} = - (+) 1,
\label{urb-alpha-CP}
\end{equation}
\begin{equation}
|e_{\pm}\rangle     \longrightarrow |K_{1(2)}\rangle
\rho_{1(2)}\Big(\,|{\bf 1}\rangle - \alpha_{1(2)}|{\bf 2}\rangle
\Big) \equiv 2^{-1/2}\Big(\,|{\bf 1}\rangle     - (+) |{\bf
2}\rangle \Big) , \label{urb-K1(2)-CP}
\end{equation}
\begin{equation}
{\zeta}_{-(+)} \longrightarrow m_{1(2)}^{LOY} = h_{0}^{LOY} -(+)
(h_{12}^{LOY}h_{21}^{LOY})^{1/2}, \label{urb-m1(2)-CP}
\end{equation}
and this is the picture which one observes within the LOY theory
in the case of conserved CP symmetry.

\subsection{The case of nonconserved CP and conserved CPT.}

It is known that in 1964 it was announced that long living $K_{2}$
states exhibited a decay into two pions, forbidden in the case  of
conserved CP--symmetry \cite{cronin-1964}. These CP--violating
decays  for the $K_{0},{\overline K}_{0}$ complex are only of
order 0.1 \% of the CP--conserving decays but nevertheless  this
is  the  proof that CP--symmetry is not conserved in neutral kaons
complex.  So, the only possibility is to assume that CP--symmetry
is  violated but CPT--symmetry is conserved. This is because
within the  context of local quantum field theory, CPT
conservation is a theorem.

Assuming  that  CPT  is  the  symmetry  for   the   system   under
investigation  and  that  subspaces  of  neutral    kaons ${\cal
H}_{\parallel}$ and their decay  products $Q{\cal H} \equiv {\cal
H}_{\perp}$ are invariant  under CPT--transformation, i.e.,
assuming (\ref{urb-[CPT,H]}), (\ref{urb-[CPT,P]}), one easily
finds that $H_{11}\equiv  H_{22}$, and
\begin{equation}
\Theta {\Sigma}(m_{0}) {\Theta}^{-1} = {\Sigma}^{+}(m_{0}),
\label{urb-CPT-Sigma}
\end{equation}
i.e.,
\begin{equation}
\Theta  H_{LOY} {\Theta}^{-1} = H_{LOY}^{+} ,
\label{urb-CPT-H-LOY-a}
\end{equation}
which implies
\begin{equation}
{\Sigma}_{11}^{R}(m_{0}) \equiv {\Sigma}_{22}^{R}(m_{0}), \; \; \;
\; {\Sigma}_{11}^{I}(m_{0}) \equiv {\Sigma}_{22}^{I}(m_{0}),
\label{urb-Sigma-11-22}
\end{equation}
and
\begin{equation}
{\Sigma}_{21}^{R}(m_{0}) \equiv [ {\Sigma}_{12}^{R}(m_{0})
]^{\ast}, \; \; \; \; {\Sigma}_{21}^{I}(m_{0}) \equiv [
{\Sigma}_{12}^{I}(m_{0}) ]^{\ast}, \label{urb-Sigma-21-12}
\end{equation}
Therefore the obvious conclusion, exploited  widely  in the
literature, is that
\begin{equation}
M_{11}^{LOY} = M_{22}^{LOY} \stackrel{\rm def}{=} M_{0}^{LOY}, \;
\; {\Gamma}_{11}^{LOY} = {\Gamma}_{22}^{LOY} \stackrel{\rm def}{=}
{\Gamma}_{0}^{LOY}, \label{urb-M-LOY-11-22-a}
\end{equation}
or,
\begin{equation}
{\Sigma}_{11}(m_{0}) \equiv {\Sigma}_{22}(m_{0}),
\label{urb-Sigma-11=22-a}
\end{equation}
which means that
\begin{equation}
h_{11}^{LOY}    \equiv    h_{22}^{LOY}    \stackrel{\rm def}{=}
h_{0}^{LOY}, \label{urb-h-0-LOY}
\end{equation}
and
\begin{equation}
M_{21}^{LOY} = (M_{12}^{LOY})^{\ast}, \; \; {\Gamma}_{21}^{LOY} =
({\Gamma}_{12}^{LOY})^{\ast}, \label{urb-M21-12-LOY-a}
\end{equation}
in  CPT--invariant  system (\ref{urb-E-pm-E}),
(\ref{urb-[CPT,P]}). Relations (\ref{urb-M-LOY-11-22-a}) and
(\ref{urb-h-0-LOY}) are the standard result of the LOY approach
and this is the picture which one meets in the literature
\cite{LOY2}
--- \cite{Bigi} and  which  one obtains searching only for
properties of matrix elements of above obtained $H_{LOY}$.

These properties mean  that  the  effective  Hamiltonian  $H_{LOY}
\equiv H_{LOY}^{\Theta}$ ($H_{LOY}^{\Theta}$ denotes the  operator
$H_{LOY}$ when the property (\ref{urb-[CPT,H]}) occurs)  is
represented by  the following $(2 \times 2)$ matrix
\begin{equation}
H_{LOY}^{\Theta} \equiv \left(
\begin{array}{cc}
M_{0} - \frac{i}{2} {\Gamma}_{0}, & M_{12}^{LOY} -
\frac{i}{2} {\Gamma}_{12}^{LOY} \\
{M_{12}^{LOY}}^{\ast} - \frac{i}{2} {{\Gamma}_{12}^{LOY}}^{\ast} &
M_{0} - \frac{i}{2} {\Gamma}_{0}
\end{array}
\right) , \label{urb-LOY-Theta}
\end{equation}
and lead to the following form of eigenvectors  and  eigenvalues
of $H_{LOY}^{\Theta}$. From (\ref{urb-alpha-pm}) and
(\ref{urb-h-0-LOY}) it follows that
\begin{equation}
|e_{+(-)}\rangle     \longrightarrow |K_{l(s)}\rangle     =
{\rho}_{l(s)}^{LOY} \,\Big(\,|{\bf 1}\rangle  -
\alpha_{l(s)}^{LOY} |{\bf 2}\rangle \Big) \equiv
{\rho}_{l(s)}^{LOY} \,\Big(\,|{\bf 1}\rangle     + (-) a |{\bf
2}\rangle \Big) , \label{urb-Kl-Ks}
\end{equation}
where
\begin{eqnarray}
{\alpha}_{+ (-)} & \equiv & {\alpha}_{l(s)}^{LOY} \stackrel{\rm
def}{=} - (+) a \equiv - (+)
\Big( \frac{h_{21}^{LOY}}{h_{12}^{LOY}} {\Big)}^{1/2},
\label{urb-CPT-alpha} \\
|a| & \neq & 1,  \label{a-neq-1}
\end{eqnarray}
and
\begin{equation}
|{\rho}_{l(s)}^{LOY}|^{2} = ( 1 + |a|^{2}\,)^{-1}.
\label{urb-rho-ls}
\end{equation}
We have also that
\begin{equation}
{\zeta}_{+(-)} \longrightarrow {\mu}_{l(s)}^{LOY} = h_{0}^{LOY} +
( - ) h^{LOY}, \label{urb-mu-ls-LOY}
\end{equation}
\begin{equation}
h^{LOY} = (h_{12}^{LOY}h_{21}^{LOY})^{1/2}. \label{urb-h-LOY-CPT}
\end{equation}
Let us notice that in this case
\begin{equation}
\langle K_{s}|K_{l}\rangle     \neq 0, \label{urb-Kl-Ks-neq-0}
\end{equation}
in contradistinction to the case of  conserved  CP--symmetry,
where  $\langle K_{1}|K_{2}\rangle    $ \linebreak  $=0$
(\ref{urb-K-12-a}). States $|K_{l}\rangle    $,  $|K_{s}\rangle $
are long and  short living superpositions   of   $K_{0}$   and
${\overline K}_{0}$ respectively. Experimentally determined
life-times are ${\tau}_{l} \simeq 5.17 \times 10^{-8}$ s,
${\tau}_{s} \simeq 0.89 \times  10^{-10}$  s,  which  mean that
$c{\tau}_{l} \simeq 15.49$ m and $c{\tau}_{s} \simeq 2.68$ cm.
There is $A(K \rightarrow 2 \pi ) \neq - A({\overline  K}
\rightarrow  2 \pi )$ in this case and therefore two-- and
three--pion decays are allowed for $K_{l}$ and $K_{s}$ both.

\subsection{The case of nonconserved CPT.}

If to assume that CPT--symmetry is not conserved in  the  physical
system under consideration, i.e., that
\begin{equation}
[ \Theta , H] \neq 0, \label{urb-[CPT,H]-neq-0}
\end{equation}
then   $h_{11}^{LOY}   \neq   h_{22}^{LOY}$,   which   imply that
$|{\alpha}_{l}| \neq |{\alpha}_{s}|$  in  expansions
(\ref{urb-e-pm-1}), (\ref{urb-ls-12}). It  is convenient to
express difference between $H_{LOY}^{\Theta}$ and the effective
Hamiltonian $H_{LOY}$ appearing  within  the LOY approach in the
case of nonconserved CPT--symmetry as follows
\begin{eqnarray}
H_{LOY} & \equiv & H_{LOY}^{\Theta} + \delta H_{LOY}
\label{urb-H-LOY+delta} \\
& = & \left(
\begin{array}{cc}
( M_{0} + \frac{1}{2} \delta M) - \frac{i}{2} ( {\Gamma}_{0} +
\frac{1}{2} \delta \Gamma ),
& M_{12}^{LOY} - \frac{i}{2} {\Gamma}_{12}^{LOY} \\
{M_{12}^{LOY}}^{\ast} - \frac{i}{2} {{\Gamma}_{12}^{LOY}}^{\ast} &
(M_{0} - \frac{1}{2} \delta M) - \frac{i}{2} ({\Gamma}_{0} -
\frac{1}{2} \delta \Gamma )
\end{array}
\right) .  \nonumber
\end{eqnarray}
Within the LOY model the $\delta M$ and $\delta \Gamma$ terms
violate CPT--symmetry.

Generally, when $[{\cal C}{\cal P},H] \neq 0$, $[\Theta  ,H]  \neq
0$, the eigenvectors $|l\rangle , |s\rangle $ of  $H_{\parallel}$
for the eigenvalues $\mu_{l(s)}$ differ from $|K_{1(2)}\rangle $
(\ref{urb-K1(2)-CP}) and $|K_{l(s)}\rangle $ (\ref{urb-Kl-Ks}),
and are not orthogonal \cite{cronin} --- \cite{Bigi},
\cite{cronin-1964} --- \cite{gell-mann}.

It is convenient to express the CP-- and CPT--violation parameters
in relation to   the   orthogonal   eigenvectors   $|K_{1}\rangle
$   and $|K_{2}\rangle $ of the ${\cal C}{\cal P}$--transformation
for the eigenvalues $\mp 1$ (\ref{urb-K1(2)-CP}). Vectors
$|l\rangle $, $|s\rangle $ written down in the $|K_{1}\rangle $,
$|K_{2}\rangle $ basis have the form (\ref{urb-ls-K12})
\cite{dafne,comins,barmin}. Usually, instead of the parameters
$\varepsilon_{l}$, $\varepsilon_{s}$, the parameters,
$\varepsilon$, (\ref{urb-eps}) and $\delta$, (\ref{urb-delta}),
are used \cite{cronin,dafne,barmin,lavoura}.

The interpretation of $\delta$ as the CPT--violating parameter
follows directly from a formula (\ref{urb-delta-D}) for this
parameter, when one inserts into this formula matrix elements
$h_{jk}^{LOY}$ of the effective Hamiltonian $H_{\parallel} \equiv
H_{LOY}$: see (\ref{urb-h-jk-LOY}), (\ref{urb-ksi-pm}). The
relation (\ref{urb-h-0-LOY}) leads to the conclusion
\begin{equation}
\delta \cong {\delta}^{LOY} = 0, \label{urb-delta-LOY=0}
\end{equation}
where  ${\delta}^{LOY}     \equiv     {\delta}(h_{jk} \equiv
h_{jk}^{LOY})$.

The parameter $\delta$ (or equivalent parameters) is usually
measured in experimental tests of CPT invariance in the neutral
kaon system \cite{Gaillard} --- \cite{cronin},
\cite{dafne,barmin,data}. The standard interpretation of the
result of this experiment (which is a straightforward consequence
of (\ref{urb-delta-LOY=0})) is that the result the result $\delta
\cong  0$ means that the property $[\Theta,H] = 0$ holds in the
investigated system, while the opposite result, $\delta \neq 0$,
means that the CPT---symmetry is not conserved in this system --
this again follows from the traditional interpretation which may
be based on (\ref{urb-[CPT,H]}), (\ref{urb-delta-LOY=0}). Indeed,
relations (\ref{urb-delta-LOY=0}) and (\ref{urb-h-0-LOY}) mean
that (see (\ref{urb-delta}))
\begin{equation}
{\varepsilon}_{l} - {\varepsilon}_{s} = 0. \label{LOY-e-l=s}
\end{equation}
Therefore the tests based on the relation (\ref{urb-Re-h11-h22})
are considered as the test of CPT--invariance and the results of
such tests are interpreted that the masses of the particle "1"
(the $K_{0}$ meson) and its antiparticle "2" (the
${\overline{K}}_{0}$ meson) must be equal if CPT--symmetry holds.
Parameters $\varepsilon_{l}, \varepsilon_{s}, \gamma_{l},
\gamma_{s}$, etc., appearing in the right side of the relation
(\ref{urb-Re-h11-h22}) can be extracted from experiments in such
tests and then these parameters can be used to estimate the left
side of this relation. The estimation for the mass difference
obtained in this way with the use of the recent data \cite{data}
reads
\begin{equation}
\frac{|M_{1} - M_{2}|}{m_{K_{0}}} = \frac{|m_{K_{0}} -
m_{{\overline{K}}_{0}}|}{m_{K_{0}}} \leq 10^{-18}, \label{mk-mk}
\end{equation}
and this estimation is considered as indicating no
 CPT--violation effect. This interpretation follows
from the properties of the $H_{LOY}$.

Summing up, according to the above, physicists believe that
\begin{eqnarray}
\delta = 0 \Leftrightarrow |M_{1} - M_{2}| = 0 \; \; &\Rightarrow
& \; \; [\Theta,H] =0\,
(?),\label{urb-delta=0} \\
\delta \neq 0 \Leftrightarrow |M_{1} - M_{2}| \neq 0 \; \; &
\Rightarrow & \; \; [\Theta,H] \neq 0, (?).
\label{urb-delta-neq-0}
\end{eqnarray}

This  is  the standard result of the LOY approach and this is the
picture  which one meets in  the literature \cite{LOY1}  ---
\cite{Bigi}.

\section{Real properties  of  time  evolution \\ in  subspace  of
neutral kaons.}

The aim of this Section is to show that the diagonal matrix
elements of the exact effective Hamiltonian $H_{||}$ can not be
equal when the total system under consideration is CPT invariant
but CP noninvariant.

Universal properties of the (unstable) particle--antiparticle
subsystem of the system described by the  Hamiltonian $H$, for
which  the relation (\ref{urb-[CPT,H]}) holds, can be extracted
from the matrix elements of the exact $U_{||}(t)$ appearing in
(\ref{urb-U||-psi}). Such $U_{||}(t)$ has the following form
\begin{equation}
U_{||}(t) = P U(t)P, \label{ku-U||}
\end{equation}
where $P$ is given by the formula (\ref{urb-P}) and $U(t)$ is the
total unitary evolution operator, which solves the
Schr\"{o}\-din\-ger equation (\ref{urb-Schrod}). Operator
$U_{||}(t)$ acts in the subspace of unstable states ${\cal H}_{||}
\equiv P {\cal H}$. Of course, $U_{||}(t)$ has nontrivial form
only if
\begin{equation}
[P, H] \neq 0, \label{ku-[P,H]}
\end{equation}
and only then transitions of states from ${\cal H}_{||}$ into
${\cal H}_{\perp}$ and vice versa, i.e., decay and regeneration
processes, are allowed.

Using the matrix representation one finds
\begin{equation}
U_{||}(t) \equiv \left(
\begin{array}{cc}
{\rm \bf A}(t) & {\rm \bf 0} \\
{\rm \bf 0} & {\rm \bf 0}
\end{array} \right)
\label{A(t)}
\end{equation}
where ${\rm \bf 0}$ denotes the suitable zero submatrices and a
submatrix ${\rm \bf A}(t)$ is the $2 \times 2$ matrix acting in
${\cal H}_{||}$
\begin{equation}
{\rm \bf A}(t) = \left(
\begin{array}{cc}
A_{11}(t) & A_{12}(t) \\
A_{21}(t) & A_{22}(t)
\end{array} \right) \label{A(t)=}
\end{equation}
and $A_{jk}(t)$ is given by (\ref{urb-A-jk}) for $|\psi\rangle =
|{\bf k}\rangle$, $(j,k =1,2)$.

Now assuming (\ref{urb-[CPT,H]}) and using, e.g., the phase
convention defined by the formula (\ref{urb-CPT1=2}) one easily
finds that \cite{chiu} --- \cite{leonid-fp}, \cite{nowakowski}
\begin{equation}
A_{11}(t) = A_{22}(t). \label{A11=A22}
\end{equation}
Note that assumptions (\ref{urb-[CPT,H]}) and (\ref{urb-CPT1=2})
give no relations between $A_{12}(t)$ and $A_{21}(t)$.

The important relation between amplitudes $A_{12}(t)$ and
$A_{21}(t)$ follows from the famous Khalfin's Theorem \cite{chiu}
--- \cite{leonid-fp}, \cite{kabir}. This Theorem states that
in the case of
unstable states, if amplitudes $A_{12}(t)$ and $A_{21}(t)$ have
the same time dependence
\begin{equation}
r(t) \stackrel{\rm def}{=} \frac{A_{12}(t)}{A_{21}(t)} = {\rm
const} \equiv r, \label{r=const},
\end{equation}
then it must be $|r| = 1$.

For unstable particles the relation (\ref{A11=A22}) means that
decay laws
\begin{equation}
p_{j}(t) \stackrel{\rm def}{=} |A_{jj}(t)|^{2}, \label{p-j}
\end{equation}
(where $j = 1,2$), of the particle $|{\bf 1}\rangle $ and its
antiparticle $|{\bf 2}\rangle $ are equal,
\begin{equation}
p_{1}(t) \equiv p_{2}(t). \label{p1=p2}
\end{equation}
The consequence of this last  property is that the decay rates of
the particle $|{\bf 1}\rangle $ and its antiparticle $|{\bf
2}\rangle $
\[{\gamma}_{j}(t) \stackrel{\rm def}{=} -
\frac{1}{p_{j}(t)} \frac{\partial p_{j}(t)}{\partial t},\] must be
equal too,
\begin{equation}
{\gamma}_{1}(t) = {\gamma}_{2}(t), \label{g1=g2}
\end{equation}
On the other hand from (\ref{A11=A22}) it does not follow that the
masses of the particle "1" and the antiparticle "2" should be
equal.

More conclusions about the properties of the matrix elements of
$H_{||}$, that is in particular about $M_{jj}$, one can infer
analyzing the following identity \cite{horwitz}, \cite{bull} ---
\cite{pra}
\begin{equation}
H_{||} \equiv H_{||}(t) = i \frac{\partial U_{||}(t)}{\partial t}
[U_{||}(t)]^{-1}, \label{H||2a}
\end{equation}
where $[U_{||}(t)]^{-1}$ is defined as follows
\begin{equation}
U_{||}(t) \, [U_{||}(t)]^{-1} = [U_{||}(t)]^{-1} \, U_{||}(t) \, =
\, P. \label{U^-1}
\end{equation}
(Note that the identity (\ref{H||2a}) holds, independent of
whether $[P,H] \neq 0$ or $[P,H]=0$). The expression (\ref{H||2a})
can be rewritten using the matrix ${\bf A}(t)$

\begin{equation}
H_{||}(t) \equiv  i \frac{\partial {\bf A}(t)}{\partial t} [{\bf
A}(t)]^{-1}. \label{H||2b}
\end{equation}
Relations (\ref{H||2a}), (\ref{H||2b}) must be fulfilled by the
exact as well as by every approximate effective Hamiltonian
governing the time evolution in every two dimensional subspace
${\cal H}_{||}$ of states $\cal H$ \cite{horwitz,bull} ---
\cite{pra}.

It is easy to find from (\ref{H||2a}) the general formulae for the
diagonal matrix elements, $h_{jj}$, of $H_{||}(t)$, in which we
are interested. We have \cite{plb-2002}
\begin{eqnarray}
h_{11}(t) &=& \frac{i}{\det {\bf A}(t)} \Big( \frac{\partial
A_{11}(t)}{\partial t} A_{22}(t) - \frac{\partial
A_{12}(t)}{\partial t} A_{21}(t) \Big), \label{h11=} \\
h_{22}(t) & = & \frac{i}{\det {\bf A}(t)} \Big( - \frac{\partial
A_{21}(t)}{\partial t} A_{12}(t) + \frac{\partial
A_{22}(t)}{\partial t} A_{11}(t) \Big). \label{h22=}
\end{eqnarray}
Now, assuming (\ref{urb-[CPT,H]}) and using the consequence
(\ref{A11=A22}) of this assumption, one finds
\begin{equation}
h_{11}(t) - h_{22}(t) =  \frac{i}{\det {\bf A}(t)} \Big(
\frac{\partial A_{21}(t)}{\partial t} A_{12}(t) - \frac{\partial
A_{12}(t)}{\partial t} A_{21}(t) \Big). \label{h11-h22=}
\end{equation}
Next, after some algebra one obtains
\begin{equation}
h_{11}(t) - h_{22}(t) = - i \, \frac{A_{12}(t) \, A_{21}(t) }{\det
{\bf A}(t)} \; \frac{\partial}{\partial t} \ln
\Big(\frac{A_{12}(t)}{A_{21}(t)} \Big). \label{h11-h22=1}
\end{equation}
This result means that in the considered case for $t>0$ the
following Theorem holds \cite{plb-2002}:\\
\hfill\\
{\bf Theorem 1.}\\
\begin{equation}
h_{11}(t) - h_{22}(t) = 0 \; \; \Leftrightarrow \; \;
\frac{A_{12}(t)}{A_{21}(t)}\;\; = \; \; {\rm const.}, \; \; (t >
0). \label{h11-h22=0}
\end{equation}
\hfill\\

Thus for $t > 0$ the problem under studies is reduced to the
Khalfin's Theorem (see the relation (\ref{r=const})).

From (\ref{h11=}) and (\ref{h22=}) it is easy to see that at $t=0$
\begin{equation}
h_{jj}(0) = \langle  {\bf j}|H|{\bf j}\rangle , \; \; (j=1,2),
\label{hjjt=0}
\end{equation}
which means that in a CPT invariant system (\ref{urb-[CPT,H]}) in
the case of pairs of unstable particles, for which transformations
of type (\ref{urb-CPT1=2}) hold
\begin{equation}
M_{11}(0) = M_{22}(0) \equiv \langle  {\bf 1}|H|{\bf 1}\rangle ,
\label{M11=M22}
\end{equation}
the unstable particles "1" and "2" are created at $t=t_{0} \equiv
0$ as  particles with equal masses. The same result can be
obtained from the formula (\ref{h11-h22=1}) by taking $t
\rightarrow 0$.

In the general case
\begin{equation}
h_{jk}(0) = H_{jk},\;\;\;(j,k=1,2). \label {e93} \\
\end{equation}

Now let us go on to analyze the  conclusions following from the
Khalfin's Theorem. CP noninvariance requires that $|r| \neq 1$
\cite{chiu,leonid1,leonid-fp,nowakowski} (see also \cite{LOY1}
--- \cite{Yu-V}, \cite{chiu} --- \cite{Bigi}). This
means that in such a case
it must be $r = r(t) \neq {\rm const.}$. So, if in the system
considered the property (\ref{urb-[CPT,H]}) holds but
\begin{equation}
[{\cal CP}, H] \neq 0, \label{[CP,H]}
\end{equation}
and the unstable states "1" and "2" are connected by a relation of
type (\ref{urb-CPT1=2}), then at $t > 0$ it must be $(h_{11}(t) -
h_{22}(t)) \neq 0$ in this system.

On the other hand to complete the discussion of the problem one
can examine consequences of the assumptions that $(h_{11}(t) -
h_{22}(t)) = 0$ is admissible for $t >0$ and that the system under
considerations need not be CP-- or CPT--invariant. In such a case
an analysis of the considerations
leading to the Theorem 1 allows one to conclude that\\
\hfill\\
\hfill\\
{\bf Conclusion 1.}\\
If $(h_{11}(t) - h_{22}(t)) = 0$ for $t>0$ then it must be\\
a)
\[ \frac{A_{11}(t)}{A_{22}(t)} = {\rm const.},\;\; {\rm and} \;\;
\frac{A_{12}(t)}{A_{21}(t)} = {\rm const.},\;\;
  {\rm for}\;\; (t > 0) ,\]
or,\\
b)
\[\frac{ A_{11}(t)}{A_{22}(t)} \neq {\rm const.},\;\; {\rm and} \;\;
\frac{A_{12}(t)}{A_{21}(t)} \neq {\rm const.},\;\;
  {\rm for}\;\; (t > 0) .\]
\hfill\\

The case a) means that CP--symmetry is conserved and there is no
any information about CPT invariance. The case b) denotes that
system under considerations is neither CP--invariant nor
CPT--invariant.

Now let us examine properties of $\Re \, (h_{11}(t) - h_{22}(t))$
for $t > t = t_{0} \equiv 0$. It can be done, e.g., using the
methods  exploited  in  \cite{plb-1993}.

In the nontrivial case (\ref{ku-[P,H]}) from (\ref{H||2a}), using
(\ref{urb-Schrod}), (\ref{urb-U(t)}) and (\ref{ku-U||}) we find
\begin{eqnarray}
H_{\parallel}(t) & \equiv &  PHU(t)P[U_{\parallel}(t)]^{-1} P
\label{e89}  \\
& \equiv & PHP + PHQ U(t) [ U_{\parallel}(t)]^{-1}P \label{r89} \\
& \stackrel{\rm def}{=} & PHP + V_{\parallel}(t).  \label{e90}
\end{eqnarray}
Thus \cite{ijmpa-1992,pla-1992}
\begin{equation}
H_{\parallel}(0)  \equiv  PHP,   \;   \; V_{\parallel}(0) = 0, \;
\; V_{\parallel} (t \rightarrow  0)  \simeq -itPHQHP,
\label{H||(0)}
\end{equation}    so,      in      general $H_{\parallel}(0)
\neq$    $H_{\parallel}(t    \gg    t_{0}=0)$ and $V_{\parallel}(t
\neq 0)  \neq  V_{\parallel}^{+}(t \neq 0)$, $H_{\parallel}(t \neq
0)  \neq  H_{\parallel}^{+}(t  \neq 0)$.

Eigenvectors  of time--dependent  $H_{\parallel}(t)$  can depend
on time $t$: From (\ref{urb-e-pm-1}) we obtain
\begin{equation}
|e_{+}\rangle  \longrightarrow |l^{t}\rangle ,
\;\;\;\;\;\;\;|e_{-}\rangle  \longrightarrow |s^{t}\rangle ,
\label{e92}
\end{equation}
and, in general, they are not orthogonal. In long time region $(t
\rightarrow \infty)$ vectors $|l(s)^{t \rightarrow \infty}\rangle
$ correspond to the states $| K_{l(s)} \rangle $ (\ref{urb-Kl-Ks})
obtained within  the LOY approach. In short time region, from
(\ref{H||(0)}) we have (\ref{e93}) and
\begin{equation}
h_{jk}(t \rightarrow 0) = H_{jk} - it\langle  {\bf j}|HQH|{\bf
k}\rangle , \label{e94}
\end{equation}
This implies that in the case considered
\begin{equation}
{\alpha}_{l(s)} (t \rightarrow 0) = -(+) \Big(
\frac{H_{21}}{H_{12}} {\Big)}^{1/2}, \label{e95}
\end{equation}
if
\begin{equation}
H_{12} \equiv \langle {\bf 1}|H|{\bf 2}\rangle = H_{21}^{\ast}
\neq 0. \label{H12-neq-0}
\end{equation}
So, in such a case
\begin{equation}
|{\rho}_{l(s)}(0)|^{2} = \frac{1}{2} . \label{e96}
\end{equation}
Therefore  the  eigenvectors  for  $H_{\parallel}(t)$   take   the
following form in the early time period:
\begin{equation}
|l(s)^{t \rightarrow 0}\rangle  = {\rho}_{l(s)}(0) \Big[ |{\bf
1}\rangle +(-) \Big( \frac{ H_{21} }{ H_{12} } {\Big)}^{1/2} |{\bf
2}\rangle  \Big] . \label{e97}
\end{equation}
They  are  orthogonal  quite independently  of  whether CP-- or
CPT--symmetries are conserved or not.

Beyond the short time region, for $t \;  >  \; 0$, the vectors
$|l(s)^{t}\rangle $  are  orthogonal  only  in  the  case   of
conserved CP--symmetry. Indeed, if property (\ref{urb-[CP,H]})
holds one has
\begin{equation}
[{\cal C}{\cal P} , U_{\parallel}(t) ] = 0, \label{e98}
\end{equation}
and thus, by the identity (\ref{e89}),
\begin{equation}
[{\cal C}{\cal P} , H_{\parallel}(t) ] = 0, \label{e99}
\end{equation}
which and definition (\ref{urb-CP-K}) lead to the following
relations
\begin{equation}
h_{11}(t) = h_{22}(t), \;\;\;\;\; \; h_{12}(t) = h_{21}(t),
\label{e100}
\end{equation}
valid for every $t$ in the considered case  of  conserved  CP.
This means that eigenvectors for $H_{\parallel}(t)$ (\ref{H||2a})
must be equal $|K_{1}\rangle $ and $|K_{2}\rangle $
(\ref{urb-K1(2)-CP}), i.e., effective Hamiltonians
$H_{\parallel}(t)$ and $H_{LOY}$ lead to the same solutions of the
eigenvalue problem in the case of conserved CP--symmetry.

Now let  us  pass  on  to  considerations  of  CPT--transformation
properties of $H_{\parallel}$. There is only one assumption for
the operator $\Theta$ describing CPT--transformation in $\cal H$:
we require the assumption (\ref{urb-[CPT,P]}) for $\Theta$ to be
fulfilled, and, in contradistinction to (\ref{urb-[CPT,H]}), there
is  not any assumptions  for $[\Theta , H]$. Using this assumption
and the identity (\ref{e89}), after some algebra, one finds
\cite{mpla-2004}
\begin{equation}
[\Theta,H_{\parallel}(t)] = {\cal A}(t) + {\cal B}(t),
\label{e102}
\end{equation}
where:
\begin{eqnarray}
{\cal A}(t) \; & = & \; P  [{\Theta},H] U(t) P \bigl(
U_{\parallel}(t)
{\bigr)}^{-1} P, \label{e103} \\
{\cal B}(t) \; & = & \;  \Big{\{}  PHQ  -  PHU(t)  P \bigl(
U_{\parallel}(t) {\bigr)}^{-1}P \Big{\}}  [{\Theta} ,U(t)] P
\bigl(
U_{\parallel}(t) {\bigr)}^{-1} P \nonumber \\
& \equiv & P \Big{\{} H \; - \; H_{\parallel}(t) P \Big{\}}
[{\Theta},U(t)] P \bigl( U_{\parallel}(t){\bigr)}^{-1} P.
\label{e104}
\end{eqnarray}

We observe that ${\cal A}(0) \equiv P[\Theta,H]P$ and ${\cal B}(0)
\equiv 0$. From definitions and general properties of operators
$\cal C$,$\cal P$ and $\cal T$ \cite{Yu-V}, \cite{messiah} ---
\cite{gibson} it is known that ${\cal T}U(t{\neq}0)$ $=$
$U_{T}^{+}(t{\neq}0){\cal T}$ $\neq$ $ U(t{\neq}0){\cal T}$
(Wigner's definition for $\cal T$ is used) \cite{wigner},  and
thereby ${\Theta}U(t \neq 0) = U_{CPT}^{+}(t \neq 0){\Theta}$
\cite{messiah,bohm,wigner}, i.e.  $[\Theta,U(t \neq  0)] \neq 0$.
So, the component ${\cal B}(t)$ in (\ref{urb-p/q}) is nonzero for
$t \neq 0$ and it is obvious that there is a chance for
$\Theta$--operator to commute with the effective  Hamiltonian
$H_{\parallel}(t  \neq 0)$   only   if $[\Theta,H] \neq 0$. On the
other hand, the property $[\Theta,H] \neq 0$ does not imply that
$[\Theta,H_{\parallel}(0)] =  0$  or $[\Theta,H_{\parallel}(0)]
\neq 0$. These two  possibilities  are admissible, but if
$[\Theta,H] =  0$  then  there  is  only  one possibility:
$[\Theta,H_{\parallel}(0)] = 0$ \cite{plb-1993}.

From (\ref{e102}) we find
\begin{equation}
\Theta H_{\parallel}(t) \Theta^{-1} - H_{\parallel}(t) \equiv
\bigl( {\cal A}(t) + {\cal B}(t) \bigr) \Theta^{-1}. \label{e105}
\end{equation}
These relations and (\ref{e89} lead to the conclusion that
generally  ${\Theta}H_{\parallel}(t{\neq}0){\Theta}^{-1}$ $\neq$
$H_{\parallel}^{+}(t{\neq}0)$, and ${\Theta}H_{\parallel}(t \neq
0){\Theta}^{-1}$ $\neq$ $H_{\parallel}(t{\neq}0)$.

Now, keeping in mind that $|{\bf 2}\rangle  \equiv
|\overline{K}_{0}\rangle $ is the antiparticle for $|{\bf
1}\rangle \equiv |K_{0}\rangle $ and that,  by definition, the
$\Theta$--operator transforms $|{\bf 1}\rangle $ in $|{\bf
2}\rangle $ (\ref{urb-CP-psi}), (the phase convention
(\ref{urb-CPT1=2}) is assumed) and taking into account another
properties of $\Theta$, we obtain from (\ref{e105})
\begin{equation}
h_{11}(t)^{\ast} - h_{22}(t) = - \langle  {\bf 2}| \bigl( {\cal
A}(t) + {\cal B}(t) \bigr) |{\bf 1}\rangle . \label{e106}
\end{equation}

Adding together expression (\ref{e106}) and its  complex conjugate
one yields

\begin{equation}
{\Re \,} \; (h_{11}(t) - h_{22}(t)) = - {\Re \,} \; \langle  {\bf
2}| \bigl( {\cal A}(t) + {\cal B}(t) \bigr) | {\bf 1}\rangle .
\label{e107}
\end{equation}
Now, let us assume for a moment that the property
(\ref{urb-[CPT,H]}) occurs, i.e., that $[\Theta,H] = 0$. Then
${\cal A}(t) \equiv 0$ and thus  $[ {\Theta}, H_{\parallel}(0) ] =
0$, which is in agreement  with  an  earlier,  similar  result
\cite{plb-1993}. In this case  we  have  ${\Theta}U(t)  =
U^{+}(t)\Theta$, which gives ${\Theta}U_{\parallel}(t) =$
$U_{\parallel}^{+}(t)\Theta$, ${\Theta}U_{\parallel}^{-1}(t)  =
(U_{\parallel}^{+}(t))^{-1}\Theta$, and
\begin{equation}
[\Theta, U(t)] = - 2i \bigl( {\Im \,} \; U(t) \bigr) \Theta.
\label{e108}
\end{equation}
This relation leads to the following  result  in  the case of the
conserved  ${\cal C}{\cal P}\cal T$-symmetry under consideration
\begin{equation}
{\cal B}(t) = -2iP \Big{\{}  H \; - \; H_{\parallel}(t) \; P
\Big{\}} \bigl( {\Im \,} \; U(t) \bigr) P \bigl(
U_{\parallel}^{+}(t) {\bigr)}^{-1}\Theta. \label{e109}
\end{equation}
From (\ref{e109}) we obtain
\begin{equation}
\langle  {\bf 2}|{\cal B}(t) |{\bf 1}\rangle  \equiv 2i \langle
{\bf 2}| \bigl( H \; - \; H_{\parallel}(t) \; P \bigr) \bigl( {\Im
\,} \; U(t) \bigr) P \bigl( U_{\parallel}^{+}(t) {\bigr)}^{-1}
|{\bf 2}\rangle . \label{e110}
\end{equation}
This expression allows us  to  conclude  that  $\langle  {\bf
2}|{\cal B}(0)|{\bf 1}\rangle  = 0$ and \linebreak ${\Re
\,}\langle  {\bf 2}|{\cal B}(t{\neq}0)|{\bf 1}\rangle $ $\neq 0$,
${\Im \,}\langle  {\bf 2}|{\cal B}(t{\neq}0)|{\bf 1}\rangle $
$\neq 0$ if condition (\ref{urb-[CPT,H]}) holds. This means that
in this case it must be $\Re \,(h_{11}(t) -
h_{22}(t)) \neq 0$ for $t \neq 0$.\\

So, there is no possibility for $\delta$ to take the value zero
for $t > 0$ in the case of conserved  CPT--symmetry in the system
considered: it must be $\delta \neq 0$ in  such  a case.

The only possibility for  $\delta$  to  equal  zero  is  if  the
nonzero contribution of ${\cal B}(t \neq 0)$ into  $(h_{11}(t)  -
h_{22}(t))$  is  compensated  by  a  nonzero   contribution   of
${\cal A}(t)$. It can be  observed  that  $\Re \,(\langle  {\bf
2}|{\cal B}(t {\neq}0)|{\bf 1}\rangle)  \neq 0$ irrespective  of
whether $\Theta$ commutes with $H$ or not, but ${\cal A}(t) \neq
0$  only appears if $[\Theta,H] \neq 0$.  So,  definition
(\ref{urb-delta-D})  of the parameter $\delta$, properties
(\ref{e102}), (\ref{e107}) and consequences of (\ref{e110}) lead
to the following conclusions for $\delta  \equiv
{\delta}(h_{jk}(t \gg t_{0} \equiv 0))$: \vspace{12pt}

{\noindent}{\bf Conclusion 2.}:\\
a) If $\delta = 0$ then it follows that $[\Theta,H]  \neq
0$, \\
b) If $[\Theta,H] = 0$ then it follows that  $\delta  \neq
0$. \\
c) If $\delta \neq 0$ then the cases $[\Theta,H] \neq  0$ or
$[\Theta,H]  = 0$ are  both possible.   \vspace{12pt}

The same conclusions are valid also when one uses a density matrix
approach  for  a  description  of  time  evolution   in $K_{0}$,
${\overline  K}_{0}$  complex  (see  \cite{Piskorski-2000}).  All
results and conclusions of this Section are, in fact, model
independent, i.e., they do not depend on the given  Hamiltonian
$H$  of  the system considered. This Section describes general
properties of two-state complex (subsystem) evolving in time in
two--dimensional subspace ${\cal H}_{\parallel}$ of the total
state  space  of the system $\cal H$ and interacting with the rest
of the system considered.

Assuming the LOY interpretation of $\Re \,(h_{jj}(t))$, ($j=1,2$),
one can conclude from the Khalfin's Theorem and from the property
(\ref{h11-h22=0}) that if $A_{12}(t), A_{21}(t) \neq 0$ for $t >
0$ and if the total system considered is CPT--invariant, but
CP--noninvariant, then $M_{11}(t) \neq M_{22}(t)$ for $t >0$, that
is, that contrary to the case of stable particles (the bound
states), the masses of the simultaneously created unstable
particle "1" and its antiparticle "2", which are connected by the
relation (\ref{urb-CPT1=2}), need not be equal  for $t
>t_{0} =0$.  Of course, such a conclusion contradicts
the standard LOY result (\ref{urb-M-LOY-11-22-a}),
(\ref{urb-h-0-LOY}). However, one should remember that the LOY
description of neutral $K$ mesons and similar complexes is only an
approximate one, and that the LOY approximation is not perfect. On
the other hand the relation (\ref{h11-h22=0}) and the Khalfin's
Theorem follow from the basic principles of the quantum theory and
are rigorous. Consequently, their implications should also be
considered rigorous.

One should remember  that  all  the  above conclusions  (as   well
as  the  conclusions  following  from  theories  based   on    the
effective Hamiltonian obtained within LOY approach) are valid   if
the experimenter is able to prepare the tested  system  such  that
its initial state fulfills condition (\ref{urb-init0}),
(\ref{urb-init}) (or (\ref{a_k(0)}) and (\ref{urb-F(0)}) ).

\section{CPT theorem and exponential decay.}

The aim of this Section is to investigate the consequences of the
main assumption of the LOY theory, (\ref{urb-LOY-main}), i. e.,
the assumption that the decay law of neutral kaons has an
exponential form. More precisely, the question if the model
assuming an exponential decay law is able to describe correctly
CPT symmetry properties of the real system will be discussed.

Assumptions of the CPT Theorem are usually formulated in terms  of
Wightman  functions $W^{(n)}$ \cite{wightman}, \cite{streater} ---
\cite{dalitz}. These  will  be considered  briefly  below (details
can be found, e.g. in \cite{streater} --- \cite{wightman}). If we
take, for example, a set of neutral scalar fields $\{
{\Phi}_{\Xi}(x) \} \ni {\phi}_{\rm A}(x)$, $ {\rm A} \in \Xi$,  $x
\equiv (ct,\vec{r}) \in M_{4}$ ($M_{4}$ is the Minkowski
space-time,   so $x^{2} \equiv c^{2}t^{2} -$ ${\vec{r}}^{2}$), ---
for the sake of simplicity only  such  fields will  be  considered
here  --- Wightman functions  are   vacuum expectation values for
products of {\em n} field functions
\begin{equation}
W_{\rm ABC \ldots}^{(n)}(x_{1},x_{2}, \ldots ,x_{n}) \equiv
\langle \Omega |{\phi}_{\rm A}(x_{1}) {\phi}_{\rm B}(x_{2})
{\phi}_{\rm C}(x_{3}) \ldots | \Omega \rangle , \label{e111}
\end{equation}
where $\Omega$  denotes  the  vacuum  state,  which  is assumed to
be unique up to a constant phase and  to  be  invariant under
transformation $U(a,{\cal A}): U(a, {\cal A})|  \Omega \rangle $
$\equiv$ $|\Omega \rangle $, where $U(a,{\cal A})$ is a continuous
unitary representation of  the  inhomogeneous  group SL(2,C): $\{
a,{\cal A} \}$ $\longrightarrow$  $U(a,{\cal A})$, $a  \in M_{4}$,
${\cal A} \in$  SL(2,C) associated  with  the restricted Lorentz
group.   The   basic requirements for quantum fields $ \{
{\Phi}_{\Xi}(x)  \}$,  which afford possibilities for writing the
Wightman functions for fields $  \{  {\Phi}_{\Xi}(x) \}$, besides
those  of  rather   strictly mathematical nature (i.e., concerning
the state space $\cal H$  of the system considered, the domain of
field operators, its  density in $\cal H$, the hermiticity, the
cyclicity of $| \Omega \rangle $,  etc. \cite{streater}
--- \cite{dalitz}) are the following: \vspace{12pt}\\
\hfill\\
{\bf (A1)} Relativistic invariance:
\begin{equation}
U(a,{\cal A}) {\phi}_{\rm A}(x) U(a,{\cal A})^{-1} = {\phi}_{\rm
A}({\cal A}x + a), \label{e112}
\end{equation}
for any $ {\rm A} \in \Xi$, \vspace{12pt}\\ \hfill\\
{\bf (A2)} The spectral  condition:  the  eigenvalues  of the
energy momentum operator ${\cal P}_{\mu}$ lie in or on the plus
(forward) light cone $V_{+}$.

The point $p = 0$ is the isolate eigenvalue of  $\cal  P$  to  the
eigenvector $| \Omega \rangle $  --- the vacuum. The operator
${\cal P}^{\mu}{\cal P}_{\mu}$ $\equiv$ $m^{2}$ is  interpreted as
the square of the mass (the assumption  (A2)  is  equivalent with
requiring all energies to be positive). We have $U(a,{\bf {\it
1}})$ $\equiv \exp (i {\cal P}^{\mu}a_{\mu})$. \vspace{12pt}\\
\hfill\\
{\bf(A3)}  Local  commutativity  (called  also "microscopic
causality"): for all pairs \linebreak ${\rm A,B} \in  \Xi$
\begin{equation}
[{\phi}_{\rm A}(x), \; {\phi}_{\rm B}(y)] = 0, \label{e113}
\end{equation}
if $(x - y)^{2} < 0, x,y \in M_{4}$.\\
\hfill\\

Having a set of fields satisfying the above  assumptions and the
vacuum state $| \Omega \rangle $, which is  cyclic,  one has a set
of uniquely defined (by the relation (\ref{e111})) Wightman
functions $ \{ W^{(n)} \}$. Conversely, a knowledge of all
Wightman functions $ \{ W^{(n)} \}$ is sufficient to characterize
a  quantum  field theory completely. Any other quantum field
theory  with  the  same values for the $ \{ W^{(n)} \}$ is
equivalent to  the  field  theory from which the original $ \{
W^{(n)} \}$  were  constructed,  up  to unitary transformation
\cite{streater} --- \cite{wightman}.

CPT--invariance could be shown to hold for  any  field  theory
which has properties listed above \cite{streater} ---
\cite{schweber}. For example,  the CPT--symmetry $\Theta$ has the
property that, for charged scalar fields ${\varphi}_{\rm A},
{\varphi}_{\rm B}$
\begin{equation}
{\Theta}^{-1} {\varphi}_{\rm A}(x_{1}) {\varphi}_{\rm B}(x_{2})
\Theta = {\varphi}_{\rm B}( - x_{2})^{+} \; \; {\varphi}_{\rm A}(
-x_{1})^{+}, \label{e114}
\end{equation}
which  can  be  written for vacuum expectation values as follows
\begin{equation}
\langle   \Omega |{\varphi}_{\rm A}(x_{1}) {\varphi}_{\rm
B}(x_{2}) | \Omega \rangle  = \langle   \Omega |{\varphi}_{\rm B}(
- x_{2})^{+} {\varphi}_{\rm B}(  -  x_{1})^{+} | \Omega \rangle .
\label{e115}
\end{equation}

For the considered case of neutral scalar fields, the requirement
of CPT--invariance in terms of Wightman  functions is written as
follows
\begin{equation}
W_{\rm AB \ldots  N}^{(n)}(x_{1},x_{2}, \ldots ,x_{n}) = W_{\rm N
\ldots  B,A}^{(n)}( - x_{n}, - x_{n-1}, \ldots , - x_{1}).
\label{e116}
\end{equation}
The essence of the CPT  Theorem  is  contained  in  the following
realization \cite{streater} --- \cite{schweber}: \vspace{12pt}

{\noindent}{\bf CPT Theorem}

If CPT--condition (\ref{e116}) holds for all $x_{1}, \ldots
,x_{n}$ then  for every $x_{1}, \ldots ,x_{n}$ such  that  $x_{1}
- x_{2}, \ldots , x_{n-1} - x_{n}$ is a   Jost  point
\cite{streater}
--- \cite{wightman},  the weak  local commutativity condition
\begin{equation}
W_{\rm AB \ldots  N}^{(n)}(x_{1},x_{2}, \ldots ,x_{n}) = W_{\rm N
\ldots  B,A}^{(n)}(x_{n},x_{n-1}, \ldots ,x_{1}) \label{e117}
\end{equation}
{\noindent}holds. Conversely, if condition (\ref{e117}) holds in a
(real) neighborhood of  a  Jost  point,  then CPT  condition
(\ref{e116})  holds everywhere. \vspace{12pt} \hfill\\

Since (A3) implies the condition (\ref{e117}),  every field theory
of  a local hermitian scalar field fulfilling  (A1)  and (A2)  has
CPT--symmetry.

Similar  theorem  is  valid  for  other  fields   fulfilling   the
conditions listed above. One should notice that in the  course  of
the proof of CPT Theorem, an  essential  role  is  played  by  the
analytical properties of $\{ W^{(n)} \}$ generated by (A2)
\cite{streater} --- \cite{dalitz}.

Having a set of quantum fields, for which CPT  Theorem  is  valid,
one can construct CPT--invariant Lagrangian density,  which  leads
to the CPT-invariant energy--momentum tensor and thus to the total
energy of the system, i.e., to the CPT--invariant Hamiltonian  $H$
of this system.

All known CPT--invariance  tests  for  neutral kaons are based  on
the LOY model, i.e., on  the  model   containing,   by  assumption
(\ref{urb-LOY-main}), exponentially decaying particles \cite{LOY1}
--- \cite{dafne}, \cite{barmin} --- \cite{Bigi}. The
problem  of the correctness of such models and
their usefulness for  designing these tests is explained by two
following theorems (see \cite{ijmpa-1998}). \vspace{12pt} \hfill\\
{\bf Theorem 2.}

If there exists $| \psi \rangle  \in \cal H$, $| \psi \rangle \neq
0$, $\langle    \psi | \psi \rangle  = 1$, where $\cal H$ is the
total Hilbert state space  of the system considered, such that
\begin{equation}
p(t;| \psi \rangle ) = |a(t)|^{2} \equiv |\langle   \psi | \exp
(-itH) | \psi \rangle  |^{2} \leq \exp (- \gamma |t|),
\label{e118}
\end{equation}
where $\gamma > 0$ and $H$ is the  total,  selfadjoint Hamiltonian
of the system  under  investigation,  then  the spectrum of $H$ is
the whole real line. \vspace{16pt} \hfill\\
{\bf Proof:} Using the projection valued  measure  $E(  \lambda
)$ for the Hamiltonian $H$:  $H| \psi \rangle $ \linebreak $= \int
\lambda {\rm d}E( \lambda  ) | \psi \rangle $, the amplitude
$a(t)$ can be expressed as follows
\begin{equation}
a(t) = \int  e^{\textstyle  -it  \lambda}  {\rm  d}
\parallel  E( \lambda ) | \psi \rangle  {\parallel}^{2}.
\label{e119}
\end{equation}
From (\ref{e118}) it follows that
\begin{equation}
\sigma ( \lambda ) \stackrel{\rm def}{=}
\parallel E( \lambda ) | \psi \rangle  {\parallel}^{2}
\label{e120}
\end{equation}
is analytically in the strip $|{\rm  Im}  \lambda  | <{\gamma}/2$.
For real $\lambda$, the function $\sigma ( \lambda )$ defines the
positive Stieltjes measure  \cite{bochner}. So,  $a(t)$ is  the
Fourier transform of this measure $\sigma ( \lambda  )$. One  can
show that $\sigma ( \lambda )$ is absolutely continuous and has  a
whole real line ${\bf R}^{1}$ as its  support  (for details  see,
e.g., \cite{sinha}), which means that the spectrum of $H$ is the
whole  real line  \cite{sinha}. (The proofs that the condition
(\ref{e118}) implies unboundedness from below the spectrum of $H$
also can be found in \cite{wiliams,vita,davies}).

Considering the case of the LOY model, one should rather use the
next theorem, which is due to Williams \cite{wiliams}: \vspace{12pt}\\
\hfill\\{\bf Theorem 3}

Let $U_{\parallel}(t) = PU(t)P$, where $U(T) = \exp (-itH)$ is a
strongly continuous, one parameter, unitary group on a Hilbert
space ${\cal H} \supset {\cal H}_{\parallel} \equiv P {\cal H}$.

(i) If
\begin{equation}
\parallel U_{\parallel}(t) |\varphi \rangle  \parallel
\leq C \exp (- \gamma t) \label{w1}
\end{equation}
for some $\gamma > 0$ and all $t > 0$, and for some nonvanishing
$| \varphi \rangle $ in ${\cal H}_{\parallel}$, then the spectrum
of $H$ is the whole real line.

(ii) The same conclusion holds if we have exponential decrease
only in the weak sense that there are two vectors $|\psi \rangle $
and $|\varphi \rangle $ in ${\cal H}_{\parallel}$ with $\langle
\psi |\varphi \rangle  \neq 0$ and
\begin{equation}
|\langle   \psi | U_{\parallel}(t)|\varphi \rangle | \leq C \exp
(- \gamma t) \label{w2}
\end{equation}
for some $\gamma >  0$ and all $t > 0$. \vspace{16pt}

The proof of this theorem can be found in \cite{wiliams}.

Conclusions following from Theorems 2 and 3 are obvious: if the
state space $\cal H$ of the system considered contains such the
state $|\psi \rangle $ that the relation (\ref{e118}) holds, or if
in the subspace unstable states ${\cal H}_{\parallel}$ of the
system under consideration are vectors $|{\bf j\rangle }$,
($j=1,2$), and $|{\psi} \rangle _{\parallel}$ such that $\langle
{\bf j}|{\psi} \rangle _{\parallel} \neq 0$ and $|\langle  {\bf
j}|U_{\parallel}(t)| {\psi} \rangle _{\parallel} | \equiv
|a_{j}(t)| \leq C \exp (-\gamma t)$ for some $\gamma > 0$ and all
$t > 0$, ($a_{j}(t)$ is defined by (\ref{urb-Schrod})), then the
CPT Theorem is not valid for these systems. This is because such
the systems do not satisfy the assumption (A2), which is necessary
for the proof of the CPT Theorem \cite{streater}
--- \cite{wightman}.

Summing up, the CPT Theorem cannot be proved for a system in which
$H$ has a spectrum equal to the whole real line, and therefore one
finds that CPT--transformation cannot be considered as the
symmetry in models containing an unstable state  $| \psi \rangle $
(or unstable states $|{\psi}_{\alpha}\rangle $)  with  assumed
exponential decay  law of type (\ref{e118}) or
(\ref{urb-LOY-main}). Simply, models with exponentially decaying
particles can not be CPT--invariant! Therefore such the models
cannot be used for studying CPT--invariance properties of physical
systems and for designing CPT--invariance tests. For designing
such the tests only such models can be used which are
CPT--invariant or which are able to be CPT--invariant and
CPT--noninvariant depending on parameters of these models. It
seems to be logic that models only CPT--noninvariant can not be
considered as appropriate for this purpose.

It seems that results of  this  Section  explain  the  difference
between predictions of the LOY model and conclusions described in
the previous Section and following from the real properties of the
system considered in the case nonconserved CPT--symmetry.

\section{An  alternative   approximation \\  for the   effective
Hamiltonian $H_{\parallel}$.}

The  aim  of  this  Section  is  to  show  that  there  exist   an
approximation consistent with general  properties  following  from
the results of Sec. 4  for the effective Hamiltonian governing
time evolution in a given subspace ${\cal H}_{\parallel}$ and to
derive an  approximate   formulae   for   this   effective
Hamiltonians $H_{\parallel} \equiv H_{\parallel}(t)$, which
CPT--transformation properties are consistent  with  those
following  from  the  real properties  of  the  system  under
consideration  ,  i.e.,  which diagonal elements, contrary to
$H_{LOY}$, are not equal  in  CPT-- invariant system
(\ref{urb-[CPT,H]}).  These  approximate formulae for
$H_{\parallel}(t)$ can be  derived using   the
Krolikowski--Rzewuski  equation   for  the projection of a state
vector \cite{KR-1,KR-2} and \cite{bull} --- \cite{pra},
\cite{Piskorski-2000}, \cite{ijmpa-1993,ijmpa-1995}. This equation
results from the Schr\"{o}dinger equation (\ref{urb-Schrod}) for
the total system under consideration \cite{KR-1,KR-2}.

So, let us consider the evolution equations for orthogonal
components $| \psi ;t\rangle_{\parallel}$  (\ref{urb-psi-P}) and
for $| \psi ; t\rangle_{\perp}$ (\ref{urb-psi-Q}) of the state
vector $| \psi ;t\rangle \equiv | \psi ;t\rangle_{\parallel} + |
\psi ; t\rangle_{\perp}$ instead of the system equations for
number functions $a_{k}(t)$, $F_{J}( \varepsilon ;t)$. Using
projection operators $P$ and $Q$, (\ref{urb-P}),  one can obtain
from the Schr\"{o}dinger equation (\ref{urb-Schrod}) for the state
vector $| \psi ;t\rangle$ two equations for its orthogonal
components $| \psi ;t\rangle_{\parallel}$ and $| \psi ;
t\rangle_{\perp}$  valid for $t \geq t_{0} = 0$:
\begin{eqnarray}
i \frac{\partial}{\partial t} | \psi ; t{\rangle_{||}} & = & PHP |
\psi ;t{\rangle_{||}} + PHQ | \psi ; t\rangle_{\perp},
\label{eq-psi-par1} \\
i \frac{\partial}{\partial t} | \psi ; t\rangle_{\perp} & = & QHQ
| \psi ;t\rangle_{\perp} + QHP | \psi ; t{\rangle_{||}},
\label{eq-psi-perp1}
\end{eqnarray}
with the initial conditions ({\ref{urb-init}), (\ref{urb-psi-par})
and (\ref{urb-F(0)}), which are equivalent to the following one
\begin{equation}
| \psi ;t = 0\rangle_{\perp} = 0. \label{psi-perp(0)}
\end{equation}

Solving Eq (\ref{eq-psi-perp1}) one can eliminate $| \psi
;t\rangle_{\perp}$ from Eq (\ref{eq-psi-par1}) by substituting the
solution of Eq (\ref{eq-psi-perp1}) back into Eq
(\ref{eq-psi-par1}). Looking for a solution of Eq
(\ref{eq-psi-perp1}) we can use the following substitution
\begin{equation}
|\widetilde{ \psi ;t} \rangle_{\perp}  \stackrel{\rm def}{=}
e^{\textstyle +itQHQ} | \psi ; t\rangle_{\perp}.  \; \; (t \geq
0), \label{psi-tilde}
\end{equation}
By means of such a substitution Eq (\ref{eq-psi-perp1}) can be
replaced by the following one
\begin{eqnarray}
i \frac{\partial}{\partial t} |\widetilde{ \psi ;t }
\rangle_{\perp}  & = & e^{\textstyle +itQHQ} QHP | \psi ;
t{\rangle_{||}},  \; \; (t \geq 0),
\label{eq-psi-tilde} \\
| \widetilde{\psi ;t} = 0\rangle_{\perp}  & = & 0. \nonumber
\end{eqnarray}
It is easy to solve this equation. Using its solution one finds
\begin{equation}
| \psi ;t\rangle_{\perp} = -i \int_{0}^{t} e^{\textstyle -i (t -
\tau ) QHQ } QHP | \psi ; \tau {\rangle_{||}} \, d \tau ,  \; \;
(t \geq 0). \label{psi-perp1}
\end{equation}

Substituting (\ref{psi-perp1}) back into Eq (\ref{eq-psi-par1})
gives for $t \geq 0$:
\begin{equation}
i \frac{\partial}{\partial t} | \psi ; t{\rangle_{||}}  = PHP |
\psi ;t{\rangle_{||}} -i \int_{0}^{t} PHQ e^{\textstyle -i (t -
\tau ) QHQ } QHP | \psi ; \tau {\rangle_{||}} \, d \tau .
\label{kr1}
\end{equation}
Notice that  Eq (\ref{kr1}) is the exact one. (In the literature,
equations of this type are called "master equation"
\cite{davies,master,Rocky}, or Krolikowski--Rzewuski equation for
the distinguished component of a state vector
\cite{acta,pra,Piskorski-2000,ijmpa-1998,ijmpa-1993,ijmpa-1995}).
It is convenient to rewrite this equation as follows
\begin{eqnarray}
( i \frac{\partial}{ {\partial} t} - PHP )|\psi ; t\rangle_{||} &
= & - i \int_{0}^{\infty} K(t - \tau ) |\psi ;  \tau \rangle_{||}
d \tau, \label{e121}  \\
|\psi ; 0\rangle_{||} & \neq & 0,\;\;\;\;\; |\psi
;0\rangle_{\perp}=0, \nonumber
\end{eqnarray}
where:
\begin{eqnarray}
K(t) & = & {\mit \Theta} (t) PHQ \exp (-itQHQ)QHP, \label{e122} \\
Q & = & {\it 1} - P, \nonumber \\
{\mit \Theta} (t) & = & { \{ } 1 \;{\rm for} \; t \geq 0, \; \; 0
\; {\rm for} \; t < 0 { \} } . \nonumber
\end{eqnarray}
Using the property (\ref{e90}) one finds from (\ref{urb-H-par}),
and (\ref{e121})
\begin{equation}
V_{\parallel} (t)|\psi ;t\rangle_{||} = - i \int_{0}^{\infty} K(t
- \tau ) |\psi ;\tau \rangle_{||}  d \tau \stackrel{\rm def}{=} -
iK \ast |\psi ;t\rangle_{||} . \label{e123}
\end{equation}
(Here the star $\ast$ denotes the convolution: $f \ast g(t) =
\int_{0}^{\infty} f(t - \tau ) g( \tau  ) \, d  \tau$). Next,
using this relation and a retarded Green's  operator  $G(t)$ for
the equation (\ref{e121})
\begin{equation}
G(t) = - i {\mit \Theta} (t) \exp (-itPHP)P, \label{e124}
\end{equation}
one obtains \cite{pra,ijmpa-1993,ijmpa-1995} for $|\psi ;
t\rangle_{||}$ having the form (\ref{urb-U||-psi})
\begin{equation}
V_{\parallel}(t) \; U_{\parallel}(t)|\psi\rangle_{||} = - i K \ast
\Big[ {\it 1} + \sum_{n = 1}^{\infty} (-i)^{n}L \ast \ldots \ast L
\Big] \ast U_{\parallel}^{(0)} (t)|\psi \rangle_{||} ,
\label{e125}
\end{equation}
where $L$ is convoluted $n$ times, ${\it 1} \equiv {\it 1}(t)
\equiv \delta (t)$,
\begin{equation}
L(t) = G \ast K(t), \label{e126}
\end{equation}
and
\begin{equation}
U_{\parallel}^{(0)} = \exp (-itPHP) \; P \label{e127}
\end{equation}
is a "free" solution of Eq.(\ref{e121}). Of course, the series
(\ref{e125}) is convergent if \linebreak $\parallel L(t)
\parallel < 1$. If for every $t \geq 0$
\begin{equation}
\parallel L(t) \parallel \ll 1,
\label{e128}
\end{equation}
then, to the lowest order of  $L(t)$,  one  finds  from
(\ref{e125}) \cite{pra,ijmpa-1993,ijmpa-1995}
\begin{equation}
V_{\parallel}(t) \cong V_{\parallel}^{(1)} (t) \stackrel{\rm
def}{=} -i \int_{0}^{\infty} K(t - \tau ) \exp {[} i ( t - \tau )
PHP {]} d \tau . \label{V||=approx}
\end{equation}
This is a general formula, valid in any ${\cal H}_{||}$, for
$V_{\parallel}(t) \cong V_{\parallel}^{(1)} (t)$ and thus (by
(\ref{H||2a}) and (\ref{e90})) for the approximate effective
Hamiltonian $H_{||} = H_{||}(t) \cong H_{||}^{(1)}(t) \equiv PHP +
V_{\parallel}^{(1)} (t)$.

To evaluate the integral (\ref{V||=approx}) it is necessary to
calculate $\exp [it PHP]$. Keeping in mind that in the case under
studies $PHP$ is the hermitian $(2 \times 2)$ matrix and using the
Pauli matrices representation ( see (\ref{urb-E-Pauli}) ),
\begin{equation}
PHP \equiv H_{0} I_{\parallel} +  \vec{H} \bullet \vec{\sigma}
\equiv H_{0}\,P +  \vec{H} \bullet \vec{\sigma}, \label{H1-pauli}
\end{equation}
where $\vec{H}$ and ${\vec{\sigma}}$ denote the following vectors:
$\vec{H} = (H_{x}, H_{y},  H_{z})$, $\vec{\sigma}$ =
(${\sigma}_{x}, {\sigma}_{y}, {\sigma}_{z}$), and the product
$\vec{H} \bullet \vec{\sigma} $ is defined analogously to
(\ref{urb-E-s}), one finds
\begin{equation}
e^{\textstyle  \pm  itPHP } = e^{\textstyle  \pm itH_{0} } \Big[
I_{||}\, \cos (t\kappa) \pm i \frac{ \vec{H} \bullet \vec{\sigma}
}{\kappa } \sin (t\kappa) \Big] ,\label{ex-H1}
\end{equation}
where
\[
H_{0}  =  \frac{1}{2} [ H_{11} + H_{22} ] ,
\]
\begin{eqnarray*}
(\kappa )^{2}  \stackrel{\rm def}{=}  \vec{H} \bullet \vec{H}& =
&  (H_{x} )^{2} + (H_{y} )^{2} +
(H_{z} )^{2}   \\
&\equiv & H_{12} H_{21} + (H_{z} )^{2} ,
\end{eqnarray*}
and
\[
H_{z}  =  \frac{1}{2} [ H_{11} - H_{22} ].
\]

It is convenient to use (\ref{H1-pauli}) again and replace
$\vec{H} \bullet\vec{\sigma}$ by $\vec{H} \bullet\vec{\sigma} = PH
P - H_{0} P$ in Eq (\ref{ex-H1}), which, after some algebra, gives
\begin{eqnarray}
e^{\textstyle  + itPHP } & \equiv & \frac{1}{2} e^{\textstyle it(
H_{0} +\kappa )} [(1 - \frac{H_{0} }{\kappa }) P + \frac{1}{\kappa
}
PHP ] \nonumber \\
& + & \frac{1}{2} e^{\textstyle it( H_{0} -\kappa )} [(1 +
\frac{H_{0} }{\kappa }) P - \frac{1}{ \kappa } PHP ] .
\label{ex-H1b}
\end{eqnarray}
This is a general form of the operator $e^{itPHP}$. Generally
$e^{itPHP}$ has the form (\ref{ex-H1b}) if $PHP$ has nonzero
off--diagonal matrix elements, ie., if the condition
(\ref{H12-neq-0}) holds.

The simplest case is the case, when
\begin{equation}
H_{12} = H_{21} = 0.  \label{H12=0}
\end{equation}
In this case one finds
\begin{equation}
PHP \equiv H_{0}\, P, \label{P-H12=0}
\end{equation}
and then that
\begin{equation}
P e^{\textstyle{ i t PHP}} = P e^{\textstyle{itH_{0}}},
\label{exp-PHP-H}
\end{equation}
and therefore the  approximate formula (\ref{V||=approx}) for
$V_{\parallel}(t)$ yields
\begin{equation}
V_{\parallel}^{(1)} (t) = - PHQ \frac{e^{\textstyle{-it(QHQ -
H_{0})}} - 1}{QHQ - H_{0}} QHP. \label{V||(t)-H0}
\end{equation}

We are rather interested in the properties of the system at long
time period, at the same for which the LOY approximation was
calculated (see [Urb-Piskorski]),  and therefore we will consider
the properties of
\begin{equation}
V_{||} \stackrel{\rm def}{=}\lim_{t \rightarrow \infty}
V_{||}^{(1)}(t). \label{V||-infty}
\end{equation}
instead of the general case $V_{||}(t) \cong V_{||}^{(1)}(t)$.

At long time period, the relation (\ref{V||(t)-H0}) leads to
\begin{equation}
V_{||} = - \Sigma (H_{0}). \label{V-H-0}
\end{equation}
This means that in the case (\ref{P-H12=0})
\begin{equation}
H_{||} = H_{0} \, P - \,\Sigma (H_{0}), \label{H||-H12=0}
\end{equation}
where the following definition was used,
\begin{equation}
H_{||} \stackrel{\rm def}{=} \lim_{t \rightarrow \infty} H_{||}(t)
\equiv PHP + \lim_{t \rightarrow \infty}V_{||}(t),
\label{H||-infty}
\end{equation}
So, in the case of $H$ such that the condition (\ref{H12=0})
occurs, one obtains the effective Hamiltonian $H_{||}$ which is
exactly the same as in the LOY approach, $H_{||} \equiv H_{LOY}$,
(compare (\ref{urb-H-LOY-op})). This means that in such a case
simply $(h_{11} - h_{22}) = 0$ when CPT symmetry is conserved.

In the general case (\ref{H12-neq-0}) inserting (\ref{ex-H1b})
into (\ref{V||=approx}) and then taking the limit
(\ref{V||-infty}) yields \cite{Urb-Pisk-2000,Piskorski-2000}
\begin{eqnarray}
V_{\parallel} & = & - \frac{1}{2} \Sigma (H_{0} + \kappa ) \Big[
(1 - \frac{H_{0}}{\kappa} )P +
\frac{1}{\kappa} PH P \Big] \nonumber \\
& \; & - \frac{1}{2} \Sigma (H_{0} - \kappa ) \Big[ (1 +
\frac{H_{0}}{\kappa} )P - \frac{1}{\kappa} PH P \Big] .
\label{V||=approx2}
\end{eqnarray}
This approximate formula for $V_{\parallel}$ leads  to  the
expressions for the matrix elements $v_{jk}(t)$, of $V_{||}$, and
$v_{jk} = \lim_{t \rightarrow \infty} v_{jk}(t)$,
\begin{equation}
v_{jk} = \langle {\bf j}|V_{||}|{\bf k}\rangle, \;\;\;\;\; (j,k
=1,2), \label{v-jk}
\end{equation}
and thus for $h_{jk}(t)$, and $h_{jk} = \lim_{t\rightarrow \infty}
h_{jk}(t)$,
\begin{equation}
h_{jk} \equiv H_{jk} + v_{jk}, \;\;\;\;\;(j,k =1,2),
\label{H+v-jk}
\end{equation}
having properties   consistent   with those following    from the
Schr\"{o}dinger equation (\ref{urb-Schrod}) and from general,
universally valid, relations  (\ref{H||2a}), (\ref{e89}) derived
in Sec. 4. The difference between these formulae and those
obtained within the LOY approach is especially visible if one
considers the matrix elements $v_{jk}(t \rightarrow \infty ) =
v_{jk}$  of $V_{\parallel}(t  \rightarrow \infty ) = V_{\parallel}
\cong V_{\parallel}^{(1)} ( \infty )$
\cite{ijmpa-1993,ijmpa-1995}, which can be used for practical
calculations of  CP-- and CPT--violation parameters in  neutral
kaons  complex. Without requiring conservation  CP-- or
CPT--symmetry  of  the system considered, the following
expressions for $v_{jk}$ can be obtained \cite{ijmpa-1993}
\begin{eqnarray}
v_{j1} = & - & \frac{1}{2} \Big( 1 + \frac{H_{z}}{\kappa} \Big)
{\Sigma}_{j1} (H_{0} + \kappa ) - \frac{1}{2} \Big( 1 -
\frac{H_{z}}{\kappa} \Big)
{\Sigma}_{j1} (H_{0} - \kappa )  \nonumber  \\
& - & \frac{H_{21}}{2 \kappa} {\Sigma}_{j2} (H_{0} + \kappa ) +
\frac{H_{21}}{2 \kappa} {\Sigma}_{j2} (H_{0} - \kappa ) ,
\label{e130}
\end{eqnarray}

\begin{eqnarray}
v_{j2} = & - & \frac{1}{2} \Big( 1 - \frac{H_{z}}{\kappa} \Big)
{\Sigma}_{j2} (H_{0} + \kappa ) - \frac{1}{2} \Big( 1 +
\frac{H_{z}}{\kappa} \Big)
{\Sigma}_{j2} (H_{0} - \kappa )  \nonumber  \\
& - & \frac{H_{12}}{2 \kappa} {\Sigma}_{j1} (H_{0} + \kappa ) +
\frac{H_{12}}{2 \kappa} {\Sigma}_{j1} (H_{0} - \kappa ) ,
\label{e131}
\end{eqnarray}
where $j,k = 1,2$. Note that these formulae for  $v_{jk}$  and
thus for $h_{jk}$ have been derived without assuming any
symmetries of a type CP--, T--, or CPT--symmetry  for  the  total
Hamiltonian H of   the system considered. It should also be
emphasized that all components  of the expressions (\ref{e130}),
(\ref{e131}) have the same  order  with  respect  to $\Sigma (
\varepsilon )$.

If to assume that the condition (\ref{urb-[CPT,H]}) holds in the
system under consideration, that is that CPT symmetry is
conserved, then in the general case (\ref{H12-neq-0}) one finds
\begin{equation}
H_{11}  = H_{22} \equiv H_{0}, \label{H11=H22}
\end{equation}
which implies that
\begin{eqnarray}
\kappa &\equiv& |H_{12} |, \label{kappa=H12}\\
H_{z} &\equiv& 0, \label{Hz=0}\\
{\Sigma}_{11}  ( \varepsilon  = {\varepsilon}^{\ast} ) &\equiv&
{\Sigma}_{22}   ( \varepsilon = {\varepsilon}^{\ast} )
\stackrel{\rm   def}{=} {\Sigma}_{0} ( \varepsilon =
{\varepsilon}^{\ast} ) . \label{Sigma11=22}
\end{eqnarray}
These relations caused by CPT symmetry of the system lead to the
following expression for $V_{||}^{\mit\Theta} = lim_{t\rightarrow
\infty} V_{||}^{\mit\Theta} (t)$, (where $V_{\parallel}^{\mit
\Theta}(t)$ denotes $V_{\parallel}(t)$ when (\ref{urb-[CPT,H]})
occurs),
\begin{eqnarray}
V_{||}^{\mit\Theta} & = & - \frac{1}{2} \Sigma (H_{0} +
|H_{12}|)\, \Big[ \Big( 1 - \frac{H_{0}}{|H{_{12}|}} \Big)P +
\frac{1}{|H_{12}|} PHP \Big]
\nonumber \\
& &  - \frac{1}{2} \Sigma (H_{0} - |H_{12}|)\, \Big[ \Big( 1 +
\frac{H_{0}}{|H{_{12}|}} \Big)P - \frac{1}{|H_{12}|} PHP \Big].
\label{V-H12n0}
\end{eqnarray}
The matrix  elements   $v_{jk}^{\mit  \Theta}$   of operator
$V_{\parallel}^{\mit \Theta}$ have the  following form
\begin{eqnarray}
v_{j1}^{\mit\Theta} = & - & \frac{1}{2} {\Big\{ } {\Sigma}_{j1}
(H_{0} + | H_{12} |)
+ {\Sigma}_{j1} (H_{0} - | H_{12} |)  \nonumber  \\
& + & \frac{H_{21}}{|H_{12}|} {\Sigma}_{j2} (H_{0} + | H_{12} |) -
\frac{H_{21}}{|H_{12}|} {\Sigma}_{j2} (H_{0} - | H_{12} |) {\Big\}
} ,\label{v-jk-theta}
\\
v_{j2}^{\mit\Theta} = & - & \frac{1}{2} {\Big\{ } {\Sigma}_{j2}
(H_{0} + | H_{12} |)
+ {\Sigma}_{j2} (H_{0} - | H_{12} |)  \nonumber  \\
& + & \frac{H_{12}}{|H_{12}|} {\Sigma}_{j1} (H_{0} + | H_{12} |) -
\frac{H_{12}}{|H_{12}|} {\Sigma}_{j1} (H_{0} - | H_{12} |) {\Big\}
}.\label{v-jk-theta+}
\end{eqnarray}

The form of these matrix elements is rather inconvenient, e.g.,
for searching for their properties depending on the matrix
elements $H_{12}$ of $PHP$. It can be done relatively simply
assuming \cite{ijmpa-1998,ijmpa-1995}
\begin{equation}
|H_{12}| \ll |H_{0} | . \label{H12<H0}
\end{equation}
Within such an assumption one finds \cite{ijmpa-1995,ijmpa-1998}
\begin{equation}
v_{j1}^{\mit\Theta} \simeq - {\Sigma}_{j1} (H_{0} ) - H_{21}
\frac{
\partial {\Sigma}_{j2} (x) }{\partial x}
\begin{array}[t]{l} \vline \, \\ \vline \,
{\scriptstyle x = H_{0} } \end{array} , \label{vj1-H12<}
\end{equation}
\begin{equation}
v_{j2}^{\mit\Theta} \simeq - {\Sigma}_{j2} (H_{0} ) - H_{12}
\frac{
\partial {\Sigma}_{j1} (x) }{\partial x}
\begin{array}[t]{l} \vline \, \\ \vline \,
{\scriptstyle x = H_{0} } \end{array} , \label{vj2-H12<}
\end{equation}
where $j = 1,2$.  One  should  stress  that  due  to  the presence
of resonance terms, derivatives $\frac{\partial}{\partial x}
{\Sigma}_{jk} (x)$ need not  be  small,  and  so  the  products
$H_{jk} \frac{\partial}{\partial x} {\Sigma}_{jk}  (x)$  in
(\ref{vj1-H12<}), (\ref{vj2-H12<}).

From this formulae we  conclude  that in CPT invariant system,
e.g., the difference between the diagonal   matrix   elements
which plays an important role in designing tests of type
(\ref{urb-Re-h11-h22}) for  the neutral kaons system, equals to
the lowest order of $|H_{12}|$ \cite{hep-ph-0202253},
\begin{equation}
h_{11}^{\mit\Theta} - h_{22}^{\mit\Theta} \simeq H_{12} \frac{
\partial {\Sigma}_{21} (x) } {\partial x}
\begin{array}[t]{l} \vline \, \\ \vline \,
{\scriptstyle x = H_{0} } \end{array} - H_{21} \frac{ \partial
{\Sigma}_{12} (x) }{\partial x}
\begin{array}[t]{l} \vline \, \\ \vline \,
{\scriptstyle x = H_{0} }\end{array} \equiv
2h_{z}^{\mit\Theta}\neq 0. \label{h11-h22-H12<}
\end{equation}
So, in a general case, in contradiction to the property
(\ref{urb-h-0-LOY}) obtained within the LOY theory, one finds for
diagonal matrix elements  of $H_{\parallel}$ calculated within the
above described approximation  that  in CPT--invariant system the
nonzero matrix elements, $H_{12} \neq 0$, of $PHP$ cause that
\begin{equation}
(h_{11}^{\mit\Theta} - h_{22}^{\mit\Theta}) \neq 0. \label{b10}
\end{equation}
(If CPT symmetry holds , $2h_{z}^{\mit \Theta} = 0$ only if
$[{\cal C}{\cal P}, H] = 0$).

Comparing relations (\ref{v-jk-theta}), (\ref{v-jk-theta+}) and
(\ref{urb-h-jk-LOY}) one can relate matrix elements,
$h_{jk}^{\mit\Theta}$ of the more accurate effective Hamiltonian,
$H_{||}$, considered in this Section, to the corresponding matrix
elements $h_{jk}^{LOY}$ of $H_{LOY}$. So, assuming that
(\ref{H12<H0})) holds  one finds in the CPT--invariant system
\begin{eqnarray}
h_{j1}^{\mit  \Theta} &\simeq& h_{j1}^{\rm LOY} -  H_{21}  \frac{
\partial  {\Sigma}_{j2} (x) }{\partial x}
\begin{array}[t]{l} \vline \, \\ \vline \, {\scriptstyle x = H_{0} }
\end{array}  ,  \label{h-a} \\  h_{j2}^{\mit   \Theta}  &\simeq&
h_{j2}^{\rm  LOY}  -  H_{12}  \frac{  \partial   {\Sigma}_{j1}
(x) }{\partial  x}  \begin{array}[t]{l}   \vline   \,   \\
\vline   \, {\scriptstyle x = H_{0} } \end{array} , \label{h-b}
\end{eqnarray}
where  $j  = 1,2$.

Eigenvectors, $|l(s)\rangle$, of this $H_{||}^{\mit\Theta}$ for
the eigenvalues $\mu_{l(s)}$, (see (\ref{urb-mu-ls}) and
(\ref{urb-ls-12})) differ from the corresponding eigenvectors of
$H_{LOY}^{\mit\Theta}$, (see (\ref{urb-LOY-Theta}) ---
(\ref{urb-mu-ls-LOY}) ). In the case considered one finds (see
(\ref{urb-alpha-ls}))
\begin{equation}
\alpha_{l} + \alpha_{s} = \frac{h_{11}^{\mit\Theta} -
h_{22}^{\mit\Theta}}{h_{12}^{\mit\Theta}} \neq 0, \label{a-l+a-s}
\end{equation}
and
\begin{equation}
|\alpha_{l}| \neq |\alpha_{s}|, \label{a-l-neq-a-s}
\end{equation}
whereas for $H_{LOY}^{\mit\Theta}$ one has (see
(\ref{urb-CPT-alpha}))
\[
\alpha_{l}^{LOY} + \alpha_{s}^{LOY} \equiv 0, \;\;\;\; {\rm
and}\;\;\;\; |\alpha_{l}^{LOY}| = |\alpha_{s}^{LOY}|.\]

Analogously with (\ref{h-a}) and (\ref{h-b}), to the lowest order
of $|H_{12} |$,  for eigenvalues ${\mu}_{l}, {\mu}_{s}$
(\ref{urb-mu-ls}) of $H_{\parallel}$, we obtain \cite{ijmpa-1995}
\begin{eqnarray}  {\mu}_{s}^{\mit  \Theta}  \simeq
{\mu}_{s}^{LOY} - \frac{1}{2} &  {\Big[  }  &  H_{12}  \Big(
\frac{
\partial {\Sigma}_{21} (x) }{\partial x} \begin{array}[t]{l}  \vline
\, \\ \vline \, {\scriptstyle x = H_{0} }  \end{array}  +  a
\frac{
\partial {\Sigma}_{0} (x) }{\partial x}  \begin{array}[t]{l}  \vline
\, \\ \vline \, {\scriptstyle x = H_{0} } \end{array} \Big)
\nonumber\\ & + & H_{21} \Big(  \frac{  \partial  {\Sigma}_{12}
(x) }{\partial x}
\begin{array}[t]{l} \vline \, \\  \vline \, {\scriptstyle x =  H_{0}
} \end{array} + a^{-1} \frac{ \partial {\Sigma}_{0}  (x)
}{\partial x} \begin{array}[t]{l} \vline \, \\  \vline  \,
{\scriptstyle  x  = H_{0} } \end{array} \Big)  {\Big] } ,
\nonumber \\
&& \label{mu-LOY} \\
{\mu}_{l}^{\mit \Theta} \simeq
{\mu}_{l}^{LOY} -  \frac{1}{2}  &  {\Big[  }  &   H_{12}   \Big(
\frac{
\partial
{\Sigma}_{21} (x) }{\partial x}  \begin{array}[t]{l}  \vline  \,  \\
\vline \, {\scriptstyle x = H_{0} } \end{array} - a \frac{
\partial
{\Sigma}_{0} (x) }{\partial  x}  \begin{array}[t]{l}  \vline  \,  \\
\vline \, {\scriptstyle x = H_{0} } \end{array}  \Big) \nonumber  \\
& + & H_{21}  \Big(  \frac{  \partial  {\Sigma}_{12}  (x)
}{\partial x}
\begin{array}[t]{l} \vline \, \\ \vline \, {\scriptstyle x = H_{0} }
\end{array} - a^{-1} \frac{ \partial {\Sigma}_{0} (x) }{\partial  x}
\begin{array}[t]{l} \vline \, \\ \vline \, {\scriptstyle x = H_{0} }
\end{array} \Big)  {\Big] } , \nonumber\end{eqnarray}
where $a$  is  defined as follows
\begin{equation}      a      \stackrel{\rm       def}{=}       \Big(
\frac{h_{21}^{LOY}}{h_{12}^{LOY}}   {\Big)}^{1/2}   , \label{b15}
\end{equation}
and ${\mu}_{s}^{LOY},\;  {\mu}_{l}^{LOY}$   are eigenvalues of
$H_{LOY}$ for eigenstates $|K_{s}>$   and   $|K_{l}>$ respectively
(see (\ref{urb-mu-ls-LOY}) ).

Note that the relation (\ref{h11-h22-H12<}) means means that  if
CPT--symmetry  is  conserved in the  system considered and
(\ref{H12<H0}) holds  then for  the parameter $\delta$
(\ref{urb-delta-D}) one finds that
\begin{equation} {\delta}^{\mit \Theta} \equiv \frac{ 2h_{z}^{\mit
\Theta} }{ D^{\mit \Theta} } \neq  0,  \label{b14}
\end{equation}  where  \[  D^{\mit \Theta} \simeq D^{LOY} - (
H_{12} + H_{21} a^{-1} ) (1 +  a)  \frac{
\partial {\Sigma}_{0} (x) }{\partial x}  \begin{array}[t]{l}  \vline
\, \\ \vline \, {\scriptstyle x =  H_{0}  }  \end{array}  ,  \]
and $D^{LOY}  =  h_{12}^{LOY}  +  h_{21}^{LOY}  +  \Delta
{\mu}^{LOY}$, $\Delta  {\mu}^{LOY}   \stackrel{\rm   def}{=}
{\mu}_{s}^{LOY}   - {\mu}_{l}^{LOY}$.

Many test of fundamental symmetries in neutral kaon and similar
complexes make use of relations (\ref{urb-Re-M12}) ---
(\ref{urb-Im-h11-h22}) and the results of such tests are
interpreted within properties of matrix elements of the LOY
effective Hamiltonian. Analysis of properties of matrix elements
of the exact effective Hamiltonian performed in Sec. 4 and
relations (\ref{h11-h22-H12<})
--- (\ref{b15}) shows that such an interpretation need not be
correct and it need not reflect the real properties of the Nature.
As an example let us analyze the test based on the relation
(\ref{urb-Re-h11-h22}). The standard interpretation of such a test
as the CPT invariance test follows from properties
(\ref{urb-h-0-LOY}) and (\ref{urb-delta-LOY=0}) of the LOY
approximation. So, let us consider in details the relation
(\ref{urb-Re-h11-h22}) using matrix elements $h_{jk}$ of the more
accurate  effective Hamiltonian. From the formula
(\ref{h11-h22-H12<}) it follows that the left side of the relation
(\ref{urb-Re-h11-h22}) takes the following form in the case of
very weak interactions allowing for the nonzero first order
transitions $|{\bf 1}> \rightleftharpoons |{\bf 2}>$, that is in
the case  when $H_{12} \neq 0$ and the property (\ref{H12<H0})
holds,
\begin{equation}
M_{1} - M_{2} \, = \, \Re \, (h_{11}^{\mit\Theta} -
h_{22}^{\mit\Theta}) = 2\, \Im \, \Big( H_{21} { \frac{\partial
{\Sigma}_{12}^{I}(x)}{\partial x} \, \vrule \,}_{x=H_{0}} \, \Big)
\, + \, \ldots  \, \neq \, 0. \label{Re-h11-h22}
\end{equation}
(Note that as a matter of fact assuming  (\ref{H12-neq-0}) one has
$H_{21} \equiv <{\bf 2}|H^{(1)}|{\bf 1}>$ in (\ref{Re-h11-h22})).
Thus taking into account this result and the implications of the
assumptions (\ref{P-H12=0}), (\ref{H12=0}) one can conclude that
\cite{hep-ph-0202253}
\begin{equation}
\Re \, (h_{11}^{\mit\Theta} - h_{22}^{\mit\Theta})\, = 0 \,
\Leftrightarrow \, |H_{12}| \, = \, 0 \label{Re=0}
\end{equation}
within the considered approximation. Finally, using result
(\ref{Re-h11-h22}) one can replace the relation
(\ref{urb-Re-h11-h22}) by the following one:
\begin{eqnarray}
2\, \Im \, \Big( <{\bf 2}|H^{I}|{\bf 1}> { \frac{\partial
{\Sigma}_{12}^{I}(x)}{\partial x} \, \vrule \,}_{x=H_{0}} \, \Big)
& \simeq & - ({\gamma}_{s} - {\gamma}_{l}) [ \tan {\phi}_{SW}\,
\Re \, (\frac{{\varepsilon}_{s}
 - {\varepsilon}_{l}}{2})   \nonumber \\
& \; & - \Im \,(\frac{{\varepsilon}_{s} - {\varepsilon}_{l}}{2})
\, ], \label{test-a}\\ & \equiv & - ({\gamma}_{s} - {\gamma}_{l})
[
\tan {\phi}_{SW}\, \Re \, (\delta)   \nonumber \\
& \; & - \Im \,(\delta) \, ], \label{test-a1}
\end{eqnarray}
Note that this relation was derived within the more accurate
approximation discussed in this Section assuming CPT invariance of
the total system under consideration.

At the end of this Section it should be pointed out that the more
accurate than the LOY approximation (\ref{V||=approx}) considered
in this Section and leading to the formulae (\ref{e130}),
(\ref{e131}) for matrix elements of the more accurate effective
Hamiltonian, $H_{||}$, is self-consistent and well defined.

\section{Final remarks}

Discussing properties of the LOY model, it should be noticed that
the assumption (\ref{urb-LOY-main})  does  not  only change the
integration properties of Eq. (\ref{urb-Eq-F(t)}) essentially
making  the integration of this equation easier, but also it
essentially changes the analytical properties of the amplitude
$F_{j}( \omega ,t)$  and thus amplitudes $a_{k}(t)$ leading to
definite properties of matrix elements of $H_{LOY}$. According  to
the Theorem 3 of Sec. 5, these changes of analytical properties of
amplitudes considered cause the LOY model to behave  like  a
system with the Hamiltonian unbounded from  below. Therefore  CPT
transformation cannot be the symmetry for such an obtained  model,
even if it was the symmetry for the  initial system.  This  means
that solutions of  Eq. (\ref{urb-Schrod}),  or Eqs.
(\ref{urb-Eq-a(t)}), (\ref{urb-Eq-F(t)}) and solutions of Eq.
(\ref{urb-LOY-Eq}) should  have  different CPT--transformation
properties. (This reservation need not take place in  the  case of
the  CP-- transformation).

On the other hand, from basic principles  of  Quantum  Theory,  it
follows that exponential decay fails at short  and  at  long  time
regions \cite{leonid-1968,leonid-1957}. It seems that this and
results of Sec. 5 mentioned explain  why  the  assumption
(\ref{urb-LOY-main})  on exponentiality  of  decay amplitudes of
$K_{0}$, ${\overline K}_{o}$ mesons is wrong if  one consider such
tiny   effects as   possible   violations   of CPT--symmetry. In
the light of the conclusions following  from Theorems 2 and 3 of
Sec. 5 and from above remarks,  it  should  be clear that relation
(\ref{urb-h-0-LOY}) $h_{11}^{LOY} = h_{22}^{LOY}$ need not be true
for real CPT--invariant systems. What is more, in view of the
conclusions obtained in Sec. 4, the conventional, standard
interpretation of  the difference of these matrix elements appears
to be wrong.

Similar reservations  concern  formulae  for the matrix elements
of $H_{eff}$ obtained within the so--called pole approximation
\cite{Bil-Kab}, which  is,  in  fact,  equivalent  to the  LOY
approach.   This approximation is obtained by replacing the
function  (the  matrix ${\Sigma}(\varepsilon )$) which appears in
the  denominator  under the integral defining the amplitudes for
which we  are  searching, and which corresponds to a real system,
by its value at  the  pole $m_{0}$ (i.e., one replaces
${\Sigma}(\varepsilon )$ by ${\Sigma}(m_{0})$ there). This
substitution  completely  changes the analytical properties  of
this  function  and  therefore  the approximate model obtained is
no longer able to describe all, perhaps, very tiny effects (such
as  possible violations of CPT--symmetry) occurring in the real
system.  The same concerns models of neutral kaons decay in which
Bell--Steinberger unitarity relation is assumed to be valid: the
exponential  decay laws  for states $|K_{0}\rangle $ and
$|{\overline  K}_{0}\rangle $  was assumed in  the course of the
derivation of this relation \cite{bell}.

Taking into account conclusions of Sec. 4, the  experimental
result $\delta \propto (h_{11} -  h_{22})$\linebreak $\neq 0$
means nothing: in such a case the  relations $[ \Theta ,H] = 0$
and $[ \Theta ,H] \neq 0$ are admissible in the system under
consideration. It is clear that it will be essential  for  the
result  of experimental tests of  the CPT--invariance $\delta = 0$
to be exact and only such a result can be understood independently
of the model. In contradistinction    to    the    standard,
conventional interpretation  \cite{LOY2}  ---  \cite{Yu-V},
\cite{barmin} --- \cite{Bigi} such a result  will prove  that
$[\Theta,H] \neq 0$ in neutral kaons, or other similar, systems.
The problem is whether the experimenter will be able to  perform
their experiments with the accuracy guaranteeing  the  proper
answer to the question of whether $\delta = 0$ or  $\delta  \neq
0$. There is a chance for the tested system that $[\Theta,H] = 0$
only if the experiment confirms the  existence  of  a  positive
small (maybe very, very small) number, say $\lambda$, such  that
$|\delta| > \lambda$. The proper interpretation  of  the  result
$\delta \neq  0$  depends on  the model  calculations  of  the
quantity $(h_{11}(t) - h_{22}(t))$, or, which is equivalent,  on
the calculated values of matrix elements  of  type  $\langle  {\bf
2}| {\cal A}(t)|{\bf 1}\rangle $ and $\langle  {\bf 2}|{\cal
B}(t)|{\bf  1}\rangle $. This can not be performed within the LOY
approach  and requires more exact approximations. It seems that
the approximation described in \cite{beyond} or, the other one,
described in Sec.6 and exploited in
\cite{ijmpa-1993,ijmpa-1995,ijmpa-1998} may be a more effective
tool for this purpose: assuming (\ref{urb-[CPT,H]}),
(\ref{urb-[CPT,P]}), $(h_{11} -h_{22}) \neq 0$ was  found in
\cite{ijmpa-1998,ijmpa-1993,ijmpa-1995} within those
approximations.

Tests consisting of a comparison of the equality  of  the  decay
laws of ${\rm K}_{0}$ and  ${\overline{\rm K}}_{0}$ mesons, i.e.
verifying the relation (\ref{p1=p2}), seem to be the only
completely  model  independent  tests for verifying the
CPT--invariance in such and similar systems.

Taking into account all the above,  it seems that all theories
describing the  time  evolution  of  the neutral kaons and similar
systems  by  means  of  the  effective Hamiltonian $H_{\parallel}$
governing their  time  evolution  in which the CPT--invariance of
$H_{\parallel}$ as a symmetry generated by  CPT---symmetry  of the
total Hamiltonian H and leading to the property
(\ref{urb-h-jk-LOY}), is assumed, are unable to give the exact
and  correct description of  all aspects of the effects connected
with  the violation  or non-- violation   of   the CP--    and,
especially, CPT--symmetries. Also, it seems that results of the
experiments with neutral kaons, etc., designed  and  carried out
on  the basis  of  expectations  of  theories  within LOY
approximation, such as tests of CPT--invariance (at least  results
of  those  in which  CPT--invariance     or CPT--noninvariance of
$H_{\parallel}$ generated by such the invariance  properties of
$H$ were essential), should be revised using other methods than
the LOY approach (e.g. using a formalism based on the Fock-Krylov
theorem \cite{krylov} and exploited in
\cite{leonid-1968,leonid-1957}, or the approach proposed  in Sec.
6).

Using the formalism briefly described in previous Section, one can
find $(h_{11} - h_{22})$ for the generalized Fridrichs--Lee model
\cite{chiu,chiu1,ijmpa-1993}. Within this toy model one finds
\cite{ijmpa-1998}, \cite{hep-ph-0202253}
\begin{eqnarray}
\Re \, (h_{11}  -  h_{22}) \stackrel{\rm df}{=} \Re \,
(h_{11}^{FL} - h_{22}^{FL}) & \simeq & i \frac{
m_{21}{\Gamma}_{12} - m_{12}{\Gamma}_{21} }{4(m_{0} - \mu )}
\nonumber \\
& \equiv & \frac{{\Im \,}(m_{12}{\Gamma}_{21})}{2(m_{0}- \mu )}.
\label{FL1}
\end{eqnarray}
This estimation has been obtained in the case of conserved
CPT--symmetry for $|m_{12}| \ll (m_{0}- \mu)$, which corresponds
to (\ref{H12<H0}). In (\ref{FL1}) ${\Gamma}_{12}, {\Gamma}_{21}$
can be identified with those appearing in  the LOY theory, $m_{0}
\equiv H_{11} = H_{22}$ can be considered  as  the kaon mass
\cite{chiu}, $m_{jk} \equiv H_{jk} \, (j,k =1,2)$, $\mu$ can be
treated as the mass of the decay products of the neutral  kaon
\cite{chiu}.

For the neutral $K$-system, to evaluate $(h_{11}^{FL} -
h_{22}^{FL})$ one can follow, e.g., \cite{dafne,chiu} and one can
take $\frac{1}{2}{\Gamma}_{21} = \frac{1}{2}{\Gamma}_{12}^{\ast}
\sim \frac{1}{2}{\Gamma}_{s} \sim 5 \times 10^{10} {\rm sec}^{-1}$
and $(m_{0} - \mu ) = m_{K} - 2m_{\pi} \sim 200$ MeV $\sim 3
\times 10^{23} {\rm sec}^{-1}$ \cite{data}. Thus
\begin{equation}
\Re \, (h_{11}  -  h_{22}) \sim \frac{{\Gamma}_{s}}{4(m_{K}-
2m_{\pi} )}\, \Im\,(H_{12}), \label{FL1a}
\end{equation}
that is,
\begin{equation}
|\Re \, (h_{11}^{FL} - h_{22}^{FL})| \sim  1,7 \times 10^{-13}
|\Im\,(m_{12})|\equiv 1,7 \times 10^{-13} |\Im\,(H_{12})| .
\label{FL2}
\end{equation}

Note that the relation (\ref{FL1a}) is equivalent to the following
one
\begin{equation}
\Re \, (h_{11}  -  h_{22}) \sim - i \frac{{\Gamma}_{s}}{4(m_{K}-
2m_{\pi} )} <{\bf 1}|H_{-}|{\bf 2}>, \label{FL1b}
\end{equation}
where $H_{-}$ is the CP odd part of the total Hamiltonian $H
\equiv H_{+} + H_{-}$. There are $H_{-} \stackrel{\rm def}{=}
\frac{1}{2}[H - ({\cal C \cal P} ) H ({\cal C \cal P} )^{+}]$ and
$H_{+} \stackrel{\rm def}{=} \frac{1}{2}[H + ({\cal C \cal P} ) H
({\cal C \cal P} )^{+}]$ (see \cite{Lee-qft,Bigi}). $H_{+}$
denotes the CP even part of $H$. We have $<{\bf 1}|H_{-}|{\bf 2}>
\equiv i  \Im \, (<{\bf 1}|H|{\bf 2}>)$\linebreak $= i\,\Im
\,(H_{12})$. According to the literature, in the case of the
superweak model for CP violation it should be $<{\bf 1}|H_{-}|{\bf
2}> \equiv i \Im\,(H_{12})$ $ \neq 0$ to the lowest order and
$<{\bf 1}|H_{-}|{\bf 2}> = 0$ in the case of a miliweak model
\cite{Lee-qft,Bigi}.

For the Fridrichs--Lee model it has been found in
\cite{ijmpa-1993} that $h_{jk}(t) \simeq h_{jk}$ practically for
$t \geq T_{as} \simeq \frac{10^{2}}{\pi (m_{0} - |m_{12}| - \mu
)}$. This leads to the following estimation of $T_{as}$ for the
neutral $K$--system: $T_{as} \sim 10^{-22}$ sec.

Dividing both sides of (\ref{FL2}) by $m_{0}$ one arrives at the
relation corresponding to (\ref{mk-mk}):
\begin{equation}
\frac{|\Re \, (h_{11}^{FL} - h_{22}^{FL})|}{m_{0}} \sim  1,7
\times 10^{-13} \frac{|\Im\,(m_{12})|}{m_{0}}\equiv 1,7 \times
10^{-13} \frac{|\Im\,(H_{12})|}{m_{0}}. \label{FL3}
\end{equation}
So, if we suppose for a moment that the result (\ref{mk-mk}) is
the only experimental result for neutral $K$ complex then it is
sufficient for $\frac{|\Im\,(H_{12})|}{m_{0}}$ to be
$\frac{|\Im\,(H_{12})|}{m_{0}} < 10^{-5}$ in order to fulfill the
estimation (\ref{mk-mk}). Of course this could be considered as
the upper bound for a possible value of the ratio
$\frac{|\Im\,(H_{12})|}{m_{0}}$ only if there were no other
experiments and no other data for the $K_{0}, {\overline{K}}_{0}$
complex. Note that form such a point o view the suitable order of
$\frac{|\Im\,(H_{12})|}{m_{0}}$ is easily reached by the
hypothetical Wolfenstein superweak interactions
\cite{Jarlskog,Wolfenstein}, which admits first order $|\Delta S|
= 2$ transitions $K_{0} \rightleftharpoons {\overline K}_{0}$,
that is, which assumes a non--vanishing first order transition
matrix $H_{12} = <{\bf 1}|H^{I}|{\bf 2}> \sim g G_{F} \neq 0$ with
$ g \ll G_{F}$. The more realistic estimation for
$\frac{|\Im\,(H_{12})|}{m_{0}}$ can be found using the property
$\frac{|\Im\,(H_{12})|}{m_{0}} \equiv \frac{|<{\bf 1}|H_{-}|{\bf
2}>|}{m_{0}}$. One can assume that $\frac{|<{\bf 1}|H_{-}|{\bf
2}>|}{m_{0}} \sim \frac{H_{-}}{H_{strong}}$. There is
$\frac{H_{-}}{H_{strong}} \sim 10^{-14} |\varepsilon |$ for the
case of the hypothetical superweak interactions (see
\cite{Lee-qft}, formula (15.138)) and thus
$\frac{|\Im\,(H_{12})|}{m_{0}} \sim 10^{-14} |\varepsilon |$.
(Using this last estimation one should remember that it follows
from the LOY theory of neutral $K$ complex). This estimation
allows one to conclude that
\begin{equation}
\frac{|\Re \, (h_{11}^{FL} - h_{22}^{FL})|}{m_{0}} \sim  1,7
\times 10^{-27}|\varepsilon |. \label{FL4}
\end {equation}
This estimation is the estimation of type (\ref{mk-mk}) and it can
be considered as a lower bound for $\frac{|\Re \, (h_{11} -
h_{22})|}{m_{0}}$. (see also \cite{Piskorski-2003}).

Note that contrary to the approximation described in Sec. 6, the
LOY approximation, as well as the similar approximation leading to
the Bell--Strinberger unitary relations \cite{bell} are unable to
detect and correctly identify effects caused by the existence (or
absence) of the interactions for which $H_{12} \neq 0$ in the
system.

Let us analyze some important observations following from
(\ref{Re-h11-h22}), (\ref{test-a}) and from the rigorous result
obtained in \cite{plb-2002}. The non--vanishing of the right hand
side of the relation (\ref{urb-Re-h11-h22}) can not be considered
as the proof that the CPT--symmetry is violated. So, there are two
general conclusions following from (\ref{Re-h11-h22}),
(\ref{Re=0}), (\ref{test-a}) and \cite{plb-2002,mpla-2004}. The
first one: the tests based on the relation (\ref{urb-Re-h11-h22})
can not be considered as CPT--symmetry tests and this is the main
conclusion of this paper. The second one: such tests should rather
be considered as the tests for the existence of new hypothetical
(superweak (?)) interactions allowing for the first order $|
\Delta S |= 2$ transitions. Simply, the left hand side of the
relation (\ref{Re-h11-h22}) can differ from zero only if the
matrix element $<{\bf 2}|H|{\bf 1}>$ is different from zero and
thus the nonzero value of the right hand side of the relation
(\ref{test-a}) means that it should be $<{\bf 2}|H^{I}|{\bf 1}>
\neq 0$.

Note that within the LOY theory one  can also obtain nonzero first
order $|\Delta S| = 2$ transitions in Standard Model for $K_{0} -
{\overline K}_{0}$ complex \cite{Buras}. The main difference
between such an effect and the effect discussed in this paper and
connected with the relations (\ref{urb-Re-h11-h22}),
(\ref{Re-h11-h22}) --- (\ref{test-a}) is that within the LOY
theory the first order $|\Delta S|=2$ transitions can appear only
for off--diagonal matrix elements, $h_{jk}^{LOY}$, ($j \neq k)$,
of the effective Hamiltonian, $H_{LOY}$, whereas within the more
accurate approximation, discussed in the previous Section,
diagonal matrix elements, $h_{11},h_{22}$, as well as
off--diagonal matrix elements of the effective Hamiltonian
$H_{||}$ depend on $H_{12},H_{21}$. Within the LOY approach,
diagonal matrix elements of $H_{LOY}$ do not depend on
$H_{12},H_{21}$. Therefore the effect discussed in this paper is
absent in the LOY theory.

On the other hand, one should remember that the non--vanishing
right hand side of the relations (\ref{urb-Re-h11-h22}),
(\ref{test-a}) can be considered as the conclusive proof that new
interactions allowing for the first order $|\Delta S |= 2$
transitions $K_{0} \rightleftharpoons {\overline K}_{0}$ exist
only if an another experiment, based on other principles,
definitively confirms that the CPT--symmetry is not violated in
$K_{0} - {\overline K}_{0}$ system.

Unfortunately the accuracy of the today's experiments is not
sufficient to improve the estimation (\ref{mk-mk}) to the order
required by (\ref{FL3}). This especially concerns the accuracy
required by our "more realistic estimation" for
$\frac{|\Im\,(H_{12})|}{m_{0}}$. Simply it is beyond today's
experiments reach. In the light of the above estimations, keeping
in mind (\ref{Re-h11-h22}), only much more accurate tests based on
the relation (\ref{urb-Re-h11-h22}) can give the answer whether
the mentioned new hypothetical interactions exists or not.

It also seems, that above results have some meaning when attempts
to describe possible deviations from conventional quantum
mechanics are made and when possible experimental tests of such a
phenomenon and CPT--invariance in the neutral  kaons system are
considered \cite{ellis,huet}. In such a case a very important role
is played by nonzero contributions to $(h_{11} - h_{22})$
\cite{ellis,huet}: The correct description of these deviations and
experiments mentioned is impossible without taking into account
the results of this Section and the above Sections 4 --- 6. This
can not be performed within the LOY approach  and requires more
exact approximations. It seems that the approximation described
and exploited in \cite{pra,ijmpa-1993,ijmpa-1995} may be a more
effective tool for  this  purpose.

Last remark, other results \cite{ijmpa-1993,ijmpa-1995,ijmpa-1998}
obtained within the approximation described in Sec. 6  suggest
also that the form of other parameters usually used to describe
properties of $K_{0} - {\overline K}_{0}$ system is different for
the case $H_{12} \neq 0$ and for the case $H_{12} = 0$. This can
be used as the basis for designing other tests for the
hypothetical new interactions.

\end{document}